\pdfoutput=1

\documentclass[twocolumn,pdftex]{IEEEtran}

\usepackage{ifpdf}
\usepackage{stfloats}
\usepackage{float}    
\usepackage{hyperref}

\usepackage{cite}

\usepackage{graphicx}
\usepackage{subfig}
\usepackage[T1]{fontenc}
\usepackage{amsmath}
\usepackage {indentfirst}
\usepackage{amssymb}
\usepackage{color}

\usepackage{algorithm} 
\usepackage{algorithmic}
\floatname{algorithm}{Algorithm}

\allowdisplaybreaks[1]

\usepackage{fixltx2e}

\usepackage{stfloats}
\begin{document}
%
\title{Achievable Rate Maximization Pattern Design for Reconfigurable MIMO Antenna Array}
%
%
%


\author{Haonan~Wang,~\IEEEmembership{Student~Member,~IEEE},~Ang~Li,~\IEEEmembership{Senior~Member,~IEEE},\\Ya-Feng~Liu,~\IEEEmembership{Senior~Member,~IEEE},~Qibo~Qin,~Lingyang~Song,~\IEEEmembership{Fellow,~IEEE},~and~Yonghui~Li,~\IEEEmembership{Fellow,~IEEE}
\thanks{H. Wang and A. Li are with the School of Information and Communications Engineering, Faculty of Electronic and Information Engineering, Xi'an Jiaotong University, Xi'an, Shaanxi 710049, China, and are also with The State Key Laboratory of Integrated Services Networks, Xidian University, Xi'an, Shaanxi, China (e-mail: ang.li.2020@xjtu.edu.cn, whn8215858@stu.xjtu.edu.cn).}
\thanks{Y.-F. Liu is with the State Key Laboratory of Scientific and Engineering Computing, Institute of Computational Mathematics and Scientific/Engineering Computing, Academy of Mathematics and Systems Science, Chinese Academy of Sciences, Beijing 100190, China (e-mail: yafliu@lsec.cc.ac.cn).}
\thanks{Q. Qin is with Wireless Network RAN Research Department, Shanghai Huawei Technologies Co. Ltd., Shanghai 201206, China (e-mail: qinqibo1@huawei.com).}
\thanks{L. Song is with the School of Electrical Engineering and Computer Science, Peking University, Beijing 100871, China (e-mail: lingyang.song@pku.edu.cn).}
\thanks{Y. Li is with the School of Electrical and Information Engineering, The University of Sydney, Sydney, NSW 2006, Australia (e-mail: yonghui.li@sydney.edu.au).}
\thanks{The work of H. Wang and A. Li was supported in part by the Young Elite Scientists Sponsorship Program by CIC (Grant No. 2021QNRC001), in part by the National Natural Science Foundation of China under Grant 62101422, in part by the Science and Technology Program of Shaanxi Province under Grant 2021KWZ-01, and in part by the Fundamental Research Funds for the Central Universities under Grant xzy012020007. The work of Y.-F. Liu was supported in part by the National Natural Science Foundation of China (NSFC) under Grant 12022116 and Grant 12288201.}
\thanks{Part of this work has been presented in 2022 IEEE 95th Vehicular Technology Conference: (VTC2022-Spring), and 2022 IEEE 12th Sensor Array and Multichannel Signal Processing Workshop (SAM).}
}

\maketitle

\begin{abstract}
Reconfigurable multiple-input multiple-output can provide performance gains over traditional MIMO by {\color{black}reshaping the channels, i.e., introducing more channel realizations}. In this paper, we focus on the {\color{black}achievable rate} maximization pattern design for reconfigurable MIMO systems. Firstly, we introduce the matrix representation of pattern reconfigurable MIMO (PR-MIMO), based on which a pattern design problem is formulated. To further reveal the effect of the radiation pattern on the wireless channel, we consider pattern design for both the single-pattern case where the optimized radiation pattern is the same for all the antenna elements, and the multi-pattern case where different antenna elements can adopt different radiation patterns. For the single-pattern case, we show that the pattern design is equivalent to a redistribution of {\color{black}gains} among all scattering paths, and an eigenvalue optimization based solution is obtained. For the multi-pattern case, we propose a sequential optimization framework with manifold optimization and eigenvalue decomposition to obtain near-optimal solutions. Numerical results validate the superiority of PR-MIMO systems over traditional MIMO in terms of {\color{black}achievable rate}, and also show the effectiveness of the proposed solutions.
\end{abstract}


%
\IEEEpeerreviewmaketitle

\section{Introduction}

\IEEEPARstart{M}{ultiple}-input multiple-output (MIMO) techniques have received extensive research attention due to {\color{black}the additional signal processing dimension compared with} single-input single-output (SISO) systems \cite{1}. This includes point-to-point (P2P) MIMO systems where both the transmitter and the receiver are equipped with multiple antennas to improve the {\color{black}communication quality} via {\color{black}realizing spatial beamforming}. The benefits further extend to multi-user MIMO (MU-MIMO) systems, where a multi-antenna transmitter serves a set of single-antenna or multi-antenna users simultaneously for spatial multiplexing. {\color{black}However, the existing systems mainly concentrate on manipulating the transmit signals to adapt the wireless channel passively.} More recently, {\color{black}it has been shown that} employing {\color{black}reconfigurable} antennas at wireless transceivers is {\color{black}able} to {\color{black}control the transmission environment proactively}. Therefore, {\color{black}reconfigurable MIMO} techniques are seen as potential candidates for 5G and beyond communication systems \cite{3}.

Reconfigurable antennas form a special class of antennas which can be configured to operate with different frequency bands, different polarizations or radiation patterns \cite{4,5}. Among different types of reconfigurable antennas, the pattern reconfigurability which is the focus of this paper, can {\color{black}achieve power reallocation} in the signal directions so that the ability of interference suppression and energy saving can be further enhanced, while the frequency reconfigurability can reduce the interference from other wireless signals operating in the same frequency band. The polarization reconfigurability can switch between left-handed circular polarization (LHCP) and right-handed circular polarization (RHCP) to reduce the polarization mismatch and employ the polarization coding. Thanks to the development of micro-electromechanical system (MEMS) switching, semi-conductor switches, liquid metals, etc., various types of reconfigurable antennas have been designed based on variable reactive loading, parasitic tuning and material modifications \cite{6}. The applications of reconfigurable antennas in MIMO systems have been studied in some pioneering works in \cite{7,8,9}. For example in \cite{7}, the throughput gain provided by reconfigurable antennas was analyzed mathematically and it was shown that reconfiguration patterns can be seen as additional channel realizations. 
As one of the earliest realizations of {\color{black}reconfigurable MIMO}, compact parasitic arrays in the form of electronically steerable parasitic antenna radiators (ESPARs) utilized {\color{black}mutual coupling to achieve multi-stream transmission} with only one active RF chain \cite{8,9}. By employing tunable loads instead of fixed loads at each parasitic antenna element, the mutual coupling could be controlled so that different radiation patterns could be formed \cite{9}.

Pattern reconfigurable MIMO {\color{black}(PR-MIMO)} can {\color{black}reshape the power distribution} in antenna directivity at the same operating frequency, and there already existed a considerable number of methods to realize {\color{black}PR-MIMO} \cite{10}. A common method was to use a switching circuit to feed a sector array such that only one array element was executed at each timeslot \cite{11,12}. By replacing the switching circuit with a reconfigurable power divider, bidirectional and omnidirectional patterns would be obtained \cite{13,14,15}. Meanwhile, other methods for realizing pattern reconfiguration include choosing different radiation units, changing the characteristic modes of radiators, etc., \cite{16,17}. Because of the ability to increase the signal transmission distance and quality, {\color{black}PR-MIMO} has been mostly promising in the region of surveillance and tracking. For example, \cite{18} designed a high-gain pattern reconfigurable MIMO antenna array to increase the power efficiency in wireless handheld terminals. What's more, \cite{19} proposed an approach to analyze the characterization of pattern reconfigurable antennas designed for MIMO systems. It revealed that {\color{black}PR-MIMO} can redirect the signal to intended users so that the energy efficiency can be increased and the communication coverage can be extended. 

The performance benefits of PR-MIMO mainly come from {\color{black}the capability of reshaping the channels}, and the exploitation of such {\color{black}performance gains} is mainly accomplished through the effective optimal mode selection scheme. Currently, several applications of PR-MIMO have been promoted in areas such as MIMO transmission \cite{20}, target detection and tracking \cite{21}, direction of arrival (DoA) estimation \cite{22,23}. In \cite{20}, the additional efficient channels generated by {\color{black}PR-MIMO} expanded the users scheduling region, which would further improve the performance. A joint user and antenna mode selection algorithm based on determinant pairing scheduling was proposed, and a greedy low-complexity iterative selection scheme was further designed. \cite{21} discussed the application of {\color{black}PR-MIMO} in the field of target detection and tracking, and proposed a Bayesian cognitive target tracking technique, which minimized the Cramér-Rao lower bound (CRLB) of the DoA parameters through an adaptive selection of the reconfigurable antennas' modes. With respect to the DoA estimation for {\color{black}PR-MIMO}, the traditional estimation technique was combined with the mode selection scheme and an improved performance could be achieved \cite{22,23}.

Despite the above studies, there are still two challenges preventing the practical application of {\color{black}PR-MIMO}. On one hand, the mode selection mechanism brings unacceptable channel estimation overhead. The channel prediction based on the correlation among the pattern modes \cite{24} and the effective mode selection scheme based on reinforcement learning (RL) \cite{25,26,27,28,29} are the major solutions to this issue currently.
On the other hand, the physical mechanism of how the radiation pattern of {\color{black}PR-MIMO} affects the channel has not been revealed, and it is not clear how to design the radiation pattern to {\color{black}improve the channel quality actively} in {\color{black}PR-MIMO} systems, which is the focus of this paper.

In this paper, we study the {\color{black}achievable rate} maximization pattern design for {\color{black}PR-MIMO} systems. We firstly introduce the matrix representation framework of {\color{black}PR-MIMO} and further formulate a pattern design problem. In order to fully reveal the effect of the reconfigurable radiation pattern on the wireless channel, we consider both the single-pattern case where the optimized radiation pattern is the same for all the antenna elements, and the multi-pattern case where each antenna element can adopt different radiation pattern. 
The main contributions of this paper are summarized as follows:
\begin{itemize}
	\item We obtain a matrix representation formulation of the MIMO channel with pattern reconfigurability. The effect of the reconfigurable pattern is described as a pattern sampling matrix, which reveals that the pattern affects the channel by redistributing the channel gains in the direction of the scattering paths. The optimal pattern design problem aimed at maximizing the {\color{black}achievable rate} is formulated.
	\item In the single-pattern case, the requirement that all the antennas adopt the same radiation pattern results into the non-convexity of the relaxed {\color{black}achievable rate} maximization pattern design problem. In order to solve the problem with low complexity and reveal the design principle of the pattern design problem, we transform the problem and show that the pattern design is equivalent to redistributing {\color{black}gains} among different scattering paths, based on which an eigenvalue optimization based {\color{black}gain} allocation scheme is derived.
	\item For the multi-pattern case, a sequential optimization framework (SOF) is proposed as a sub-optimal scheme. In order to simplify the problem, we transform the design of the correlation modification matrix into {\color{black}a sequential optimization problem} on correlation modification vectors, where a criterion to determine their sequences to be optimized is presented. The subproblem for each vector optimization is solved via manifold optimization and eigenvalue decomposition.
	\item Numerical results clearly illustrate the {\color{black}gain} redistribution process of {\color{black}PR-MIMO} in the single-pattern case and the correlation modification process in the multi-pattern case. In addition, numerical results also demonstrate the superiority of {\color{black}PR-MIMO} with near-optimal pattern design over traditional MIMO systems in terms of {\color{black}achievable rate}. 
\end{itemize}

The rest of the paper is organized as follows. In Section II, we describe the system model of {\color{black}PR-MIMO}. In Section III, the matrix representation for {\color{black}PR-MIMO} is derived, and an optimization problem of the optimal pattern design is formulated. In Section IV, pattern design algorithms are proposed for both single-pattern and multi-pattern {\color{black}PR-MIMO}. Simulation results are provided in Section IV, and the paper is concluded in Section V with future directions.

Notations: $a$, $\boldsymbol{a}$, and $\mathbf{A}$ denote scalar, vector and matrix, respectively. $(\cdot)^{*}$, $(\cdot)^{\mathrm{T}}$, $(\cdot)^{\mathrm{H}}$ and $\operatorname{Tr}\left(\cdot\right)$ denote conjugate, transposition, conjugate transposition and trace of a matrix, respectively. $\operatorname{vec}\left(\cdot\right)$ denotes the vectorization operator and $\operatorname{diag}\left(\cdot\right)$ transforms a vector into a diagonal matrix. $\left\langle\cdot\right\rangle$ denotes the inner product of two vectors, i.e., $\left\langle\boldsymbol{a},\boldsymbol{b}\right\rangle=\boldsymbol{a}^{\mathrm{H}}\boldsymbol{b}$ for complex vectors $\boldsymbol{a}$ and $\boldsymbol{b}$. Frobenius norm of a matrix is denoted by $\|\cdot\|_{\mathrm{F}}$. $\mathbf{I}_{M}$ and $\mathbf{1}_{M \times N}$ denote a $M \times M$ identity matrix and a $M \times N$ with all entries being $1$. $\mathcal{S}_L$ is the set of all $L \times L$ symmetric matrices. $\boldsymbol{e}_i$ is the $i$-th column of an identity matrix. $\succeq$ and $\preceq$ are used as generalized inequalities for two vectors. {\color{black}$\mathbf{A} \succeq 0$ to mean that $\mathbf{A}$ is a positive semidifinite matrix and $\mathbf{A} \geqslant \boldsymbol{0}$ to mean that $\mathbf{A}$ is entry-wise nonnegative.} $\otimes$ and $\odot$ denote Kronecker and Hadamard product, respectively.
\section{System Model}

\begin{figure*}[htbp]
	\centering
	\includegraphics[width=7in]{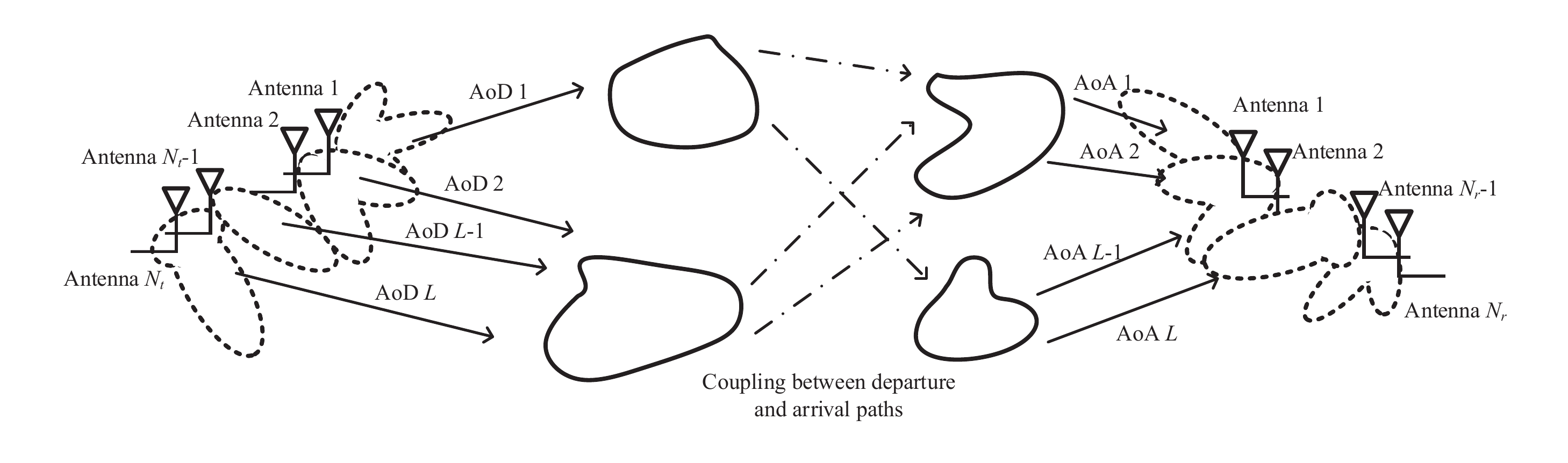}
	\caption{Multi-path channel model for {\color{black}PR-MIMO}.}
	\label{structure}
\end{figure*}
In this section, we firstly present the preliminaries of the pattern reconfigurable antenna array. Subsequently, the multi-path channel model and the channel model incorporating pattern reconfigurable antenna array are introduced, respectively.

For the pattern reconfigurable antenna array, we introduce $\vec{f}(\theta, \phi, \mu)$ to denote the complex far-field radiation pattern, where $\theta$ and $\phi$ denote the corresponding vertical and horizontal angle, and $\mu$ is the index of the antenna pattern mode. The spherical coordinate representation of the complex electric field for the $m$-th antenna element can be expressed as 
\begin{equation}
	\vec{f}\left(\theta, \phi, \mu_{m}\right)=f_{\theta}\left(\theta, \phi, \mu_{m}\right) \vec{e}_{\theta}+f_{\phi}\left(\theta, \phi, \mu_{m}\right) \vec{e}_{\phi},
\end{equation}
where $\vec{e}_{\theta}$ and $\vec{e}_{\phi}$ denote the unit vectors in the $\theta$ and $\phi$ direction, respectively.
\subsection{Multi-Path MIMO Channel Model}
We consider a single-user MIMO system in the downlink, where the number of transmit and receive antennas is $N_t$ and $N_r$ respectively, with $N_r \leq N_t$. Considering the uniform linear array (ULA) at both the BS and the receiver, the 2D physical multi-path MIMO channel model is given by \cite{7}:
\begin{equation}
	\mathbf{H}=\sum_{l=1}^{L}\alpha_{l}\boldsymbol{a}_{\mathrm{R}}\left(\theta_{l}\right)\boldsymbol{a}_{\mathrm{T}}^{\mathrm{H}}\left(\varphi_{l}\right),
\end{equation}
where $\mathbf{H} \in \mathbb{C}^{N_r \times N_t}$ is the channel matrix with $\mathbb{E}\left[\|\mathbf{H}\|_{\mathrm{F}}^{2}\right]=N_{t} N_{r}$, $L$ is the number of scattering paths with $L=N_{\mathrm{cl}}\times N_{\mathrm{ray}}$ where $N_{\mathrm{cl}}$ denotes the number of scattering clusters and $N_{\mathrm{ray}}$ denotes the number of scattering rays in each cluster, $\alpha_l$ is the complex gain of the $l$-th path. In (2), $\boldsymbol{a}_{\mathrm{R}}\left(\theta_{l}\right)$ and $\boldsymbol{a}_{\mathrm{T}}\left(\varphi_{l}\right)$ denote the receive and transmit array response vectors, where $\theta_{l}$ and $\varphi_{l}$ stand for the azimuth angles of arrival and departure (AoAs
and AoDs), respectively. The array response vectors are given by 
\begin{equation}
	\boldsymbol{a}_{\mathrm{R}}\left(\theta_{l}\right)=\dfrac{1}{\sqrt{N_r}} \left[1, e^{-j 2 \pi \frac{d_{\mathrm{R}}}{\lambda} \sin \theta_{l}}, \ldots, e^{-j 2 \pi \frac{d_{\mathrm{R}}}{\lambda}\left(N_{r}-1\right) \sin \theta_{l}}\right]^{\mathrm{T}}
\end{equation}
and
\begin{equation}
	\boldsymbol{a}_{\mathrm{T}}\left(\varphi_{l}\right)=\dfrac{1}{\sqrt{N_t}} \left[1, e^{-j 2 \pi \frac{d_{\mathrm{T}}}{\lambda} \sin \varphi_{l}}, \ldots, e^{-j 2 \pi \frac{d_{\mathrm{T}}}{\lambda}\left(N_{t}-1\right) \sin \varphi_{l}}\right]^{\mathrm{T}},
\end{equation}
where $d_{\mathrm{T}}$ and $d_{\mathrm{R}}$ denote the antenna spacing at the transmitter and the receiver, $\lambda$ is the carrier wavelength. Then $(2)$ can be further expressed in a matrix form as:
\begin{equation}
	\mathbf{H}=\mathbf{A}_{\mathrm{R}} \mathbf{\Lambda} \mathbf{A}_{\mathrm{T}}^{\mathrm{H}},
\end{equation}
where
\begin{equation}
	\mathbf{A}_{\mathrm{R}}=\left[\boldsymbol{a}_{\mathrm{R}}\left(\theta_{1}\right), \boldsymbol{a}_{\mathrm{R}}\left(\theta_{2}\right), \ldots, \boldsymbol{a}_{\mathrm{R}}\left(\theta_{L}\right)\right]
\end{equation}
and
\begin{equation}
	\mathbf{A}_{\mathrm{T}}=\left[\boldsymbol{a}_{\mathrm{T}}\left(\varphi_{1}\right), \boldsymbol{a}_{\mathrm{T}}\left(\varphi_{2}\right), \ldots, \boldsymbol{a}_{\mathrm{T}}\left(\varphi_{L}\right)\right]
\end{equation}
are array response matrices of the receiver and the transmitter, and $\mathbf{\Lambda}=\operatorname{diag}\left\{\alpha_{1}, \alpha_{2}, \ldots, \alpha_{L}\right\}$ is the complex channel gains of all scattering paths.

It should be noted that the above channel model $\mathbf{H}$ only captures the physical wireless propagation environments without considering the antenna pattern at either the transmitter or the receiver. For convenience, we call it {\it physical channel} in the following part.

\newcounter{TempEqCnt}                         
\setcounter{TempEqCnt}{\value{equation}} 
\setcounter{equation}{10}                           
\begin{figure*}[hb]
	\hrulefill
	\begin{equation}
		\begin{aligned}
			\widetilde{\mathbf{H}}= & \sum_{l=1}^{L} \alpha_{l} \boldsymbol{a}_{\mathrm{R}}\left(\theta_{l}\right) \boldsymbol{a}_{\mathrm{T}}^{\mathrm{H}}\left(\varphi_{l}\right) \odot\left({\left[\begin{array}{cl}
					F_{\mathrm{v}, \theta_{l}, v_{1}}^{\mathrm{Re}} & F_{\mathrm{h}, \vartheta_{l}, v_{1}}^{\mathrm{Re}}  \\
					F_{\mathrm{v}, \theta_{l}, v_{2}}^{\mathrm{Re}} & F_{\mathrm{h}, \vartheta_{l}, v_{2}}^{\mathrm{Re}}  \\
					\vdots & \vdots  \\
					F_{\mathrm{v}, \theta_{l}, v_{N_{r}}}^{\mathrm{Re}} & F_{\mathrm{h}, \vartheta_{l}, v_{N_{r}}}^{\mathrm{Re}}
				\end{array}\right]}  \left[\begin{array}{llll}
					F_{\mathrm{v}, \varphi_{l}, \mu_{1}}^{\mathrm{Tr}} & F_{\mathrm{v}, \varphi_{l}, \mu_{2}}^{\mathrm{Tr}} & \ldots & F_{\mathrm{v}, \varphi_{l}, \mu_{N_{t}}}^{\mathrm{Tr}} \\
					\\
					F_{\mathrm{h}, \phi_{l}, \mu_{1}}^{\mathrm{Tr}} & F_{\mathrm{h}, \phi_{l}, \mu_{2}}^{\mathrm{Tr}} & \ldots & F_{\mathrm{h}, \phi_{l}, \mu_{N_{t}}}^{\mathrm{Tr}}
				\end{array}\right]\right).
		\end{aligned}
	\end{equation}
\end{figure*}
\setcounter{equation}{\value{TempEqCnt}}

\subsection{Channel Model for {\color{black}PR-MIMO}}

Fig. 1 illustrates the {\color{black}PR-MIMO} system. Considering pattern reconfigurable antennas at both the transmitter and the receiver, the element-wise pattern reconfigurable MIMO channel model\footnote{{\color{black}Since this is the first work discussing the achievable rate maximization pattern design analytically, we focus on} continuous adjustable reconfigurable {\color{black}systems} without any physical implementation constraints to {\color{black}reveal the theoretical performance gain, which serves as a performance upper bound for reconfigurable MIMO systems with practical reconfigurable patterns.}} is represented as \cite{24}
\begin{equation}
	\begin{aligned}
		{\color{black}{\widetilde{h}}_{n, m}\left(\mu_m, v_n\right)}=\sum_{l=1}^{L} \alpha_{l} & \left\langle\vec{f}_{\mathrm{T}}\left(\varphi_{l}, \phi_{l}, \mu_m\right), \vec{f}_{\mathrm{R}}\left(\theta_{l}, \vartheta_{l}, v_n\right)\right\rangle \\
		& e^{j 2 \pi\left(\frac{d_{\mathrm{T}}}{\lambda}(m-1) \sin \varphi_{l}-\frac{d_{\mathrm{R}}}{\lambda}(n-1) \sin \theta_{l}\right)},
	\end{aligned}
\end{equation}
where {\color{black}$\widetilde{h}_{n, m}\left(\mu_m, v_n\right)$} denotes the channel between the $m$-th transmit antenna with mode $\mu_m$ and the $n$-th receive antenna with mode $v_n$, while $\mu_m \in \mathcal{R}_m$ and $v_n \in \mathcal{R}_n$ where $\mathcal{R}_m$ and $\mathcal{R}_n$ denote the continuous radiation pattern mode indication sets of the $m$-th transmit antenna and the $n$-th receive antenna, $\vec{f}_{\mathrm{T}}\left(\varphi_{l}, \phi_{l}, \mu_m\right)=\left[F_{\mathrm{v}, \varphi_{l}, \mu_m}^{\mathrm{Tr}}, F_{\mathrm{h}, \phi_{l}, \mu_m}^{\mathrm{Tr}}\right]^{\mathrm{T}}$ and $\vec{f}_{\mathrm{R}}\left(\theta_{l}, \vartheta_{l}, v_n\right)=\left[F_{\mathrm{v}, \theta_{l}, v_n}^{\mathrm{Re}}, F_{\mathrm{h}, \vartheta_{l}, v_n}^{\mathrm{Re}}\right]^{\mathrm{T}}$ are the electric-field (E-field) pattern vectors of corresponding antenna element with mode $\mu_m$ at the transmitter and that with mode $v_n$ at the receiver with $\mathrm{AoD}=\varphi_{l}$, $\mathrm{EoD}=\phi_{l}$, $\mathrm{AoA}=\theta_{l}$ and $\mathrm{EoA}=\vartheta_{l}$, {\color{black}where $\mathrm{EoD}$ and $\mathrm{EoA}$ are abbreviations for elevation angle of departure and that of arrival, respectively.} $F_{\mathrm{v}}$ and $F_{\mathrm{h}}$ denote the vertical and horizontal components of the pattern vector in spherical coordinates. Compared with $(2)$, the effect of the antenna pattern is described by the inner product of the corresponding pattern vectors. Similarly, we term $\widetilde{\mathbf{H}}$ the {\it pattern channel} matrix.

The transmission process of the {\color{black}PR-MIMO} system can be written as
\begin{equation}
	\mathbf{y}=\widetilde{\mathbf{H}}\mathbf{x}+\mathbf{n},
\end{equation}
where $\mathbf{x}$ and $\mathbf{y}$ represent transmit and receive signal vectors, and $\mathbf{n}\sim \mathcal{CN}\left(0, \sigma_{n}^{2}\mathbf{I}_{N_r}\right)$ is the additive white Gaussian noise (AWGN) vector at the receiver side.
\subsection{{\color{black}Achievable Rate of PR-MIMO}}
{\color{black}We assume the perfect channel state information at the transmitter (CSIT) and the receiver (CSIR), where the CSI at the transmitter side is used to design the radiation pattern. According to \cite{7}, the {\color{black}achievable rate} of a PR-MIMO system is given by}
\begin{equation}
	C=\log _{2} \operatorname{det}\left(\mathbf{I}_{N_{r}}+\frac{\rho}{N_{r}} \widetilde{\mathbf{H}} \widetilde{\mathbf{H}}^{\mathrm{H}}\right),
\end{equation}
where $\rho$ denotes the transmit signal-to-noise ratio (SNR).

\section{{\color{black}Achievable Rate} Maximization Pattern Design}

In this section, the matrix representation of the pattern channel is proposed, based on which the {\color{black}achievable rate} maximization pattern design is studied.

\subsection{Matrix Representation of the Pattern Channel}

Firstly, we should explain the necessity of describing the pattern channel in a matrix form. At present, the pattern channel is represented in $(8)$, which only describes the generation process of each element in the channel matrix. However, such an expression does not fully reveal how the pattern affects the channel, and meanwhile brings difficulties to the optimal pattern design.

The description in $(8)$ can be extended to a matrix form, as shown in $(11)$ at the bottom of this page. Based on that, the pattern channel matrix can be divided into vertical and horizontal components in spherical coordinates, which can be denoted as
\setcounter{equation}{11} 
\begin{equation}
	\widetilde{\mathbf{H}}=\widetilde{\mathbf{H}}_{\mathrm{v}}+\widetilde{\mathbf{H}}_{\mathrm{h}},
\end{equation}
where
\begin{equation}
	\begin{aligned}
		\widetilde{\mathbf{H}}_{\mathrm{v}} &= \sum_{l=1}^{L} \alpha_{l} \boldsymbol{a}_{\mathrm{R}}\left(\theta_{l}\right) \boldsymbol{a}_{\mathrm{T}}^{\mathrm{H}}\left(\varphi_{l}\right) \\
		& \odot\left(\left[\begin{array}{ccc}
			F_{\mathrm{v}, \varphi_{l}, \mu_{1}}^{\mathrm{Tr}} F_{\mathrm{v}, \theta_{l}, v_{1}}^{\mathrm{Re}}&  \cdots & F_{\mathrm{v}, \varphi_{l}, \mu_{N_{t}}}^{\mathrm{Tr}} F_{\mathrm{v}, \theta_{l}, v_{1}}^{\mathrm{Re}} \\
			\vdots  & \ddots & \vdots \\
			F_{\mathrm{v}, \varphi_{l}, \mu_{1}}^{\mathrm{Tr}} F_{\mathrm{v}, \theta_{l}, v_{N_r}}^{\mathrm{Re}} &  \cdots & F_{\mathrm{v}, \varphi_{l}, \mu_{N_{t}}}^{\mathrm{Tr}} F_{\mathrm{v}, \theta_{l}, v_{N_{r}}}^{\mathrm{Re}}
		\end{array}\right]\right)\\
		&= \sum_{l=1}^{L} \alpha_{l} \left(\boldsymbol{a}_{\mathrm{R}}\left(\theta_{l}\right) \odot \boldsymbol{m}_{\mathrm{v},l}^{\mathrm{Re}}\right) \left(\boldsymbol{a}_{\mathrm{T}}\left(\varphi_{l}\right) \odot \boldsymbol{m}_{\mathrm{v},l}^{\mathrm{\mathrm{Tr}}} \right)^{\mathrm{H}} \\
		&=\left(\mathbf{A}_{\mathrm{R}} \odot \mathbf{M}_{\mathrm{v}}^{\mathrm{Re}} \right) \mathbf{\Lambda}\left(\mathbf{A}_{\mathrm{T}} \odot \mathbf{M}_{\mathrm{v}}^{\mathrm{Tr}}\right)^{\mathrm{H}}
	\end{aligned}
\end{equation}
and similarly
\begin{equation}
		\widetilde{\mathbf{H}}_{\mathrm{h}} =\left(\mathbf{A}_{\mathrm{R}} \odot \mathbf{M}_{\mathrm{h}}^{\mathrm{Re}} \right) \mathbf{\Lambda}\left(\mathbf{A}_{\mathrm{T}} \odot \mathbf{M}_{\mathrm{h}}^{\mathrm{Tr}}\right)^{\mathrm{H}}.
\end{equation}
In $(13)$ and $(14)$, $\boldsymbol{m}_{\mathrm{v},l}^{\mathrm{Tr}}=\left[F_{\mathrm{v},\varphi_l,\mu_1}^{\mathrm{Tr}},F_{\mathrm{v},\varphi_l,\mu_2}^{\mathrm{Tr}},\ldots,F_{\mathrm{v},\varphi_l,\mu_{N_t}}^{\mathrm{Tr}}\right]^{\mathrm{T}}$ and $\boldsymbol{m}_{\mathrm{v},l}^{\mathrm{Re}}=\left[F_{\mathrm{v},\theta_l,v_1}^{\mathrm{Re}},F_{\mathrm{v},\theta_l,v_2}^{\mathrm{Re}},\ldots,F_{\mathrm{v},\theta_l,v_{N_r}}^{\mathrm{Re}}\right]^{\mathrm{T}}$ are the vertical pattern sampling vectors at the transmitter and the receiver in the direction of the $l$-th scattering path, similarly $\boldsymbol{m}_{\mathrm{h},l}^{\mathrm{Tr}}=\left[F_{\mathrm{h},\phi_l,\mu_1}^{\mathrm{Tr}},F_{\mathrm{h},\phi_l,\mu_2}^{\mathrm{Tr}},\ldots,F_{\mathrm{h},\phi_l,\mu_{N_t}}^{\mathrm{Tr}}\right]^{\mathrm{T}}$ and $\boldsymbol{m}_{\mathrm{h},l}^{\mathrm{Re}}=\left[F_{\mathrm{h},\vartheta_l,v_1}^{\mathrm{Re}},F_{\mathrm{h},\vartheta_l,v_2}^{\mathrm{Re}},\ldots,F_{\mathrm{h},\vartheta_l,v_{N_r}}^{\mathrm{Re}}\right]^{\mathrm{T}}$ are the horizontal ones. Combining the pattern sampling vectors together, $\mathbf{M}_{\mathrm{v}}^{\mathrm{Tr}}=\left[\boldsymbol{m}_{\mathrm{v},1}^{\mathrm{\mathrm{Tr}}},\boldsymbol{m}_{\mathrm{v},2}^{\mathrm{\mathrm{Tr}}}, \ldots , \boldsymbol{m}_{\mathrm{v},L}^{\mathrm{\mathrm{Tr}}}\right]$ and $\mathbf{M}_{\mathrm{v}}^{\mathrm{Re}}=\left[\boldsymbol{m}_{\mathrm{v},1}^{\mathrm{\mathrm{Re}}},\boldsymbol{m}_{\mathrm{v},2}^{\mathrm{\mathrm{Re}}}, \ldots, \boldsymbol{m}_{\mathrm{v},L}^{\mathrm{\mathrm{Re}}}\right]$ are the vertical pattern sampling matrices at the transmitter and the receiver, while $\mathbf{M}_{\mathrm{h}}^{\mathrm{Tr}}=\left[\boldsymbol{m}_{\mathrm{h},1}^{\mathrm{\mathrm{Tr}}},\boldsymbol{m}_{\mathrm{h},2}^{\mathrm{\mathrm{Tr}}}, \ldots, \boldsymbol{m}_{\mathrm{h},L}^{\mathrm{\mathrm{Tr}}}\right]$ and $\mathbf{M}_{\mathrm{h}}^{\mathrm{Re}}=\left[\boldsymbol{m}_{\mathrm{h},1}^{\mathrm{\mathrm{Re}}},\boldsymbol{m}_{\mathrm{h},2}^{\mathrm{\mathrm{Re}}}, \ldots, \boldsymbol{m}_{\mathrm{h},L}^{\mathrm{\mathrm{Re}}}\right]$ are the horizontal ones.

\setcounter{equation}{16}                           
\begin{figure*}[hb]
	\hrulefill
	\begin{equation}
		\begin{aligned}
			\mathcal{P}_{1}: \max _{\mathbf{M}} \ & \log _{2} \operatorname{det}\left({\mathbf{I}_{N_{r}}+\frac{\rho}{N_{r}} \mathbf{A}_{\mathrm{R}} \mathbf{\Lambda}\left(\mathbf{A}_{\mathrm{T}} \odot \mathbf{M}\right)^{\mathrm{H}}} {\left(\mathbf{A}_{\mathrm{T}} \odot \mathbf{M}\right) \mathbf{\Lambda}^{\mathrm{H}} \mathbf{A}_{\mathrm{R}}^{\mathrm{H}}}\right) \\
			\text { s.t. } & \operatorname{Tr}\left(\mathbf{A}_{\mathrm{R}} \mathbf{\Lambda}\left(\mathbf{A}_{\mathrm{T}} \odot \mathbf{M}\right)^{\mathrm{H}}\left(\mathbf{A}_{\mathrm{T}} \odot \mathbf{M}\right) \mathbf{\Lambda}^{\mathrm{H}} \mathbf{A}_{\mathrm{R}}^{\mathrm{H}}\right) \leq N_{r} N_{t}, \\
			& \mathbf{M}_{i, j} \geq 0 \quad i, j=1,2, \ldots, L.
		\end{aligned}
	\end{equation}
\end{figure*}
\setcounter{equation}{\value{TempEqCnt}}
\setcounter{TempEqCnt}{\value{equation}} 
\setcounter{equation}{20}                           
\begin{figure*}[hb]
	\hrulefill
	\begin{equation}
		\begin{aligned}
			 & -\log _{2} \operatorname{det}\left(\mathbf{I}_{N_{r}}+\frac{\rho}{N_{r}} \mathbf{A}_{\mathrm{R}} \mathbf{\Lambda}\left(\mathbf{R}_{\mathrm{T}} \odot \mathbf{X}\right) \mathbf{\Lambda}^{\mathrm{H}} \mathbf{A}_{\mathrm{R}}^{\mathrm{H}}\right) \\
			= & -\log _{2} \operatorname{det}\left(\mathbf{I}_{N_{r}}+ \frac{\rho}{N_{r}}\left[\widetilde{\boldsymbol{a}}_{\mathrm{R}, 1}, \ldots, \widetilde{\boldsymbol{a}}_{\mathrm{R}, L}\right]  \left[\begin{array}{ccc}
				x_{1,1} \boldsymbol{a}_{\mathrm{T}, 1}^{\mathrm{H}} \boldsymbol{a}_{\mathrm{T}, 1} & \cdots & x_{1, L} \boldsymbol{a}_{\mathrm{T}, 1}^{\mathrm{H}} \boldsymbol{a}_{\mathrm{T}, L} \\
				\vdots & \ddots & \vdots \\
				x_{L, 1} \boldsymbol{a}_{\mathrm{T}, L}^{\mathrm{H}} \boldsymbol{a}_{\mathrm{T}, 1} & \cdots & x_{L, L} \boldsymbol{a}_{\mathrm{T}, L}^{\mathrm{H}} \boldsymbol{a}_{\mathrm{T}, L}
			\end{array}\right]\left[\begin{array}{c}
				\widetilde{\boldsymbol{a}}_{\mathrm{R}, 1}^{\mathrm{H}} \\
				\vdots \\
				\widetilde{\boldsymbol{a}}_{\mathrm{R}, L}^{\mathrm{H}}
			\end{array}\right] \right),\\ 
			= & -\log _{2} \operatorname{det}\left(\mathbf{I}_{N_{r}}+ \frac{\rho}{N_{r}}\left[\mathbf{A}_{1}, \ldots, \mathbf{A}_{L}\right]\left[\begin{array}{ccc}
				x_{1,1} \mathbf{I}_{N_{t}} &  \cdots & x_{1, L} \mathbf{I}_{N_{t}} \\
				\vdots &  \ddots & \vdots \\
				x_{L, 1} \mathbf{I}_{N_{t}} &  \cdots & x_{L, L} \mathbf{I}_{N_{t}}
			\end{array}\right]\left[\begin{array}{c}
				\mathbf{A}_{1}^{\mathrm{H}} \\
				\vdots \\
				\mathbf{A}_{L}^{\mathrm{H}}
			\end{array}\right] \right).
		\end{aligned}
	\end{equation}
\end{figure*}

In this paper, for simplicity we {\color{black}only discuss the pattern reconfigurability at the transmitter without other reconfiguration dimensions, where the vertical polarization is considered as the constant polarization state for all conditions. I}n this case, the receiver pattern is simplified into 
\setcounter{equation}{14}
\begin{equation}
	\left[\begin{array}{cc}
		F_{\mathrm{v}, \theta_{l}, v_{1}}^{\mathrm{Re}} & F_{\mathrm{h}, \vartheta_{l}, v_{1}}^{\mathrm{Re}}  \\
		F_{\mathrm{v}, \theta_{l}, v_{2}}^{\mathrm{Re}} & F_{\mathrm{h}, \vartheta_{l}, v_{2}}^{\mathrm{Re}}  \\
		\vdots & \vdots  \\
		F_{\mathrm{v}, \theta_{l}, v_{N_{r}}}^{\mathrm{Re}} & F_{\mathrm{h}, \vartheta_{l}, v_{N_{r}}}^{\mathrm{Re}}
	\end{array}\right]=\left[\begin{array}{cc}
		1 & 0 \\
		1 & 0 \\
		\vdots & \vdots \\
		1 & 0
	\end{array}\right],
\end{equation}
based on which $(11)$ is further simplified into
\begin{equation}
    \widetilde{\mathbf{H}}=\mathbf{A}_{\mathrm{R}}\mathbf{\Lambda}\left(\mathbf{A}_{\mathrm{T}}\odot\mathbf{M}\right)^{\mathrm{H}},
\end{equation}
where $\mathbf{M}=\mathbf{M}_{\mathrm{v}}^{\mathrm{Tr}}$.

From $(16)$, it is clear that the influence of the pattern on the channel can be regarded as an additional power gain in the corresponding direction. The matrix $\mathbf{M}$ in (16) is therefore termed as the pattern sampling matrix, whose $(k,l)$-th element denotes the sampling of the $k$-th antenna element's pattern in the direction of $\mathrm{AoD}=\varphi_l$. {\color{black}When the transmit array is equipped with omni antennas, all the elements of the pattern sampling matrix $\mathbf{M}$ are equal to 1 for the radiation isotropy and the pattern channel model will reduce to the physical channel in this condition.} Moreover, $\mathbf{M}$ in $(16)$ describes the modification process of the MIMO channel by using reconfigurable antennas, which motivates us to design the pattern, which is shown in the following.

\subsection{Problem Formulation}
In this paper, we focus on the {\color{black}PR-MIMO} pattern design to improve the performance over traditional MIMO. We aim to maximize the achievable rate of the PR-MIMO system, and the optimization problem can be expressed as $\mathcal{P}_1$ in $(17)$ at the bottom of this page.
The first constraint enforces that total channel gains remain constant during the pattern design {\color{black}such that no additional power assumption will be introduced \cite{7}. Based on that, the fairness between the pattern channel and the physical channel can be ensured.} The second one constrains $\mathbf{M}$ to be a positive real matrix such that the pattern design only affects the channel path gain in the corresponding direction, without introducing additional phase adjustment overhead. Considering that each transmit antenna adopts the same radiation pattern, $\mathcal{P}_1$ can be {\color{black}re-organized as}
\setcounter{equation}{17}
\begin{equation}
	\begin{aligned}
		\mathcal{P}_{2}: \min _{\mathbf{X}} \ &-\log _{2} \operatorname{det}\left(\mathbf{I}_{N_{r}}+\frac{\rho}{N_{r}} \mathbf{A}_{\mathrm{R}} \mathbf{\Lambda}\left(\mathbf{R}_{\mathrm{T}} \odot \mathbf{X}\right) \mathbf{\Lambda}^{\mathrm{H}} \mathbf{A}_{\mathrm{R}}^{\mathrm{H}}\right) \\
		\text { s.t. } &\operatorname{Tr}\left(\mathbf{A}_{\mathrm{R}} \mathbf{\Lambda}\left(\mathbf{R}_{\mathrm{T}} \odot \mathbf{X}\right) \mathbf{\Lambda}^{\mathrm{H}} \mathbf{A}_{\mathrm{R}}^{\mathrm{H}}\right) \leq  N_{r} N_{t}, \\
		& \mathbf{X} \in \mathcal{C}_{L},\ {\color{black}\operatorname{rank}\left(\mathbf{X}\right)=1,}
	\end{aligned}
\end{equation}
where $\mathbf{X}=\mathbf{M}^{\mathrm{T}}\mathbf{M}$ is the real positive symmetric covariance matrix of $\mathbf{M}$ and $\mathcal{C}_L$ is the closed convex cone of completely positive (CP) matrices. {\color{black}$\mathbf{R}_{\mathrm{T}}=\mathbf{A}_{\mathrm{T}}^{\mathrm{H}}\mathbf{A}_{\mathrm{T}}$ is the self-covariance matrix of the transmit response matrix $\mathbf{A}_{\mathrm{T}}$.} 

Define the sets of CP matrices and doubly nonnegative (DNN) matrices as below \cite{30}
\begin{equation}
	\begin{aligned}
		\mathcal{C}_{L}:=&\left\{\mathbf{X} \in \mathcal{S}_{L}: \mathbf{X}=\mathbf{N} \mathbf{N}^{\mathrm{T}} \text { for some } \mathbf{N} \geqslant \boldsymbol{0}\right\}, \\
		\mathcal{D}_{L}:=&\left\{\mathbf{X} \in \mathcal{S}_{L}: \mathbf{X} \succeq 0, \mathbf{X} \geqslant \boldsymbol{0}\right\},
	\end{aligned}
\end{equation}
{\color{black}where $\mathbf{N}\geqslant \boldsymbol{0}$ denotes an arbitrary positive matrix and $\mathcal{S}_L$ denotes the set of all $L \times L$ symmetric matrices.}

It is obvious that each CP matrix is DNN, i.e., $\mathcal{C}_L\subseteq \mathcal{D}_L$, but the reverse is generally not true. Therefore, $\mathcal{P}_2$ can be relaxed as
\begin{equation}
	\begin{aligned}
		\mathcal{P}_{3}: \min _{\mathbf{X}}\ &-\log _{2} \operatorname{det}\left(\mathbf{I}_{N_{r}}+\frac{\rho}{N_{r}} \mathbf{A}_{\mathrm{R}} \mathbf{\Lambda}\left(\mathbf{R}_{\mathrm{T}} \odot \mathbf{X}\right) \mathbf{\Lambda}^{\mathrm{H}} \mathbf{A}_{\mathrm{R}}^{\mathrm{H}}\right) \\
		\text { s.t. } &\operatorname{Tr}\left(\mathbf{A}_{\mathrm{R}} \mathbf{\Lambda}\left(\mathbf{R}_{\mathrm{T}} \odot \mathbf{X}\right) \mathbf{\Lambda}^{\mathrm{H}} \mathbf{A}_{\mathrm{R}}^{\mathrm{H}}\right) \leq  N_{r} N_{t}, \\
		& \mathbf{X} \in \mathcal{D}_{L},\ {\color{black}\operatorname{rank}\left(\mathbf{X}\right)=1.}
	\end{aligned}
\end{equation}

The equivalence between {\color{black}$\mathcal{C}_L$} and {\color{black}$\mathcal{D}_L$} is conditional. Firstly, it is shown in \cite{30} that {\color{black}for} $L \leq 4$, $\mathcal{C}_L=\mathcal{D}_L$. Secondly, a DNN matrix $\mathbf{X}$ is also a CP matrix when $\mathrm{rank}\left(\mathbf{X}\right)=1$, which reveals that the equivalence between $\mathcal{P}_2$ and $\mathcal{P}_3$ holds in the single-pattern case, which will be discussed in Section IV.




{\color{black}Considering that the Hadamard product is not easy to handle, we further transform the cost function of $\mathcal{P}_3$,} as shown in (21) at the bottom of the previous page, where $\widetilde{\mathbf{A}}_{\mathrm{R}}=\mathbf{A}_{\mathrm{R}} \mathbf{\Lambda}=\left[\widetilde{\boldsymbol{a}}_{\mathrm{R}, 1}, \widetilde{\boldsymbol{a}}_{\mathrm{R}, 2}, \cdots, \widetilde{\boldsymbol{a}}_{\mathrm{R}, L}\right]$ is the composite receiver array response matrix which describes the power gain and the directivity of the scattering path simultaneously. $\mathbf{A}_{i}=\alpha_{i} \boldsymbol{a}_{\mathrm{R}, i} \boldsymbol{a}_{\mathrm{T}, i}^{\mathrm{H}}$ denotes the composite subchannel matrix of the $i$-th scattering path. With $\mathbf{A}_{N_{t} L \times N_{r}}=\left[\mathbf{A}_{1}^{*}, \mathbf{A}_{2}^{*}, \cdots, \mathbf{A}_{L}^{*}\right]^{\mathrm{T}}$, $\mathcal{P}_3$ can be further simplified into $\mathcal{P}_4$ as
\setcounter{equation}{21}
\begin{equation}
	\begin{aligned}
		\mathcal{P}_{4}: \min _{\mathbf{X}} \ &-\log _{2} \operatorname{det}\left(\mathbf{I}_{N_{r}}+\frac{\rho}{N_{r}} \mathbf{A}^{\mathrm{H}}\left(\mathbf{X} \otimes \mathbf{I}_{N_{t}}\right) \mathbf{A}\right) \\
		\text { s.t. } & \operatorname{Tr}\left(\mathbf{A}^{\mathrm{H}}\left(\mathbf{X} \otimes \mathbf{I}_{N_{t}}\right) \mathbf{A}\right) \leq N_{r} N_{t}, \\
		&\mathbf{X}_{i, j} \geq 0,\mathbf{X}=\mathbf{X}^{\mathrm{T}},\mathbf{X}\succeq 0.
	\end{aligned}
\end{equation}

$\mathcal{P}_4$ is equivalent to $\mathcal{P}_3$, which is a convex semidefinite programming (SDP) problem and can be solved by CVX directly. 
Although $\mathcal{P}_2$ can be solved via solving $\mathcal{P}_4$ using the semidefinite relaxation (SDR) when $\mathrm{rank}\left(\mathbf{X}\right)=1$, a more efficient solution for the single-pattern case is needed. What's more, the study of pattern design for the multi-pattern case is still blank. In what follows, we will propose the pattern designing algorithms for both the single-pattern and multi-pattern cases.

{\it Insight:} The form of $\mathcal{P}_4$ reveals that the influence of the pattern on {\color{black}achievable rate} is applied through the covariance matrix of the pattern sampling matrix. In other words, if different patterns lead to the same covariance matrix, they will achieve the same {\color{black}achievable rate}. This is a great advantage in practical pattern design, because we can choose the pattern with the lowest physical implementation cost after obtaining the optimal pattern covariance matrix.
\section{Proposed Pattern Design Algorithm}
The CP constraint of $\mathcal{P}_2$ brings difficulties to the solution of the optimization problem. Although the equivalence between $\mathcal{P}_2$ and $\mathcal{P}_4$ when $\mathrm{rank}\left(\mathbf{X}\right)=1$ makes it possible to solve $\mathcal{P}_2$ via solving $\mathcal{P}_4$ using the SDR algorithm, 
the rank-1 solution cannot be guaranteed and the principle of the pattern design problem cannot be revealed.
Thus, an efficient pattern design algorithm is needed.
Therefore in this section, we consider alternative solutions to the pattern design problem. In particular, we firstly study a simplified case where all the antennas adopt the same radiation pattern, and then extend to a more generic case where each antenna can adopt different radiation patterns.

\subsection{Single-Pattern Case}

When each antenna element employs the same radiation pattern, mathematically it is equivalent to that all the entries in each column of $\mathbf{M}$ have the same value, which makes $\mathbf{M}$ a rank-one matrix. 
With $\mathbf{M}=\mathbf{1}_{N_t \times 1}\boldsymbol{m}^{\mathrm{T}}$ where $\boldsymbol{m}=\left[m_1,m_2,\ldots,m_L\right]^{\mathrm{T}}$, (16) can be further simplified as
\setcounter{equation}{22}
\begin{equation}
	\begin{aligned}
		\widetilde{\mathbf{H}} &=\mathbf{A}_{\mathrm{R}} \mathbf{\Lambda}\left(\mathbf{A}_{\mathrm{T}} \odot \mathbf{M}\right)^{\mathrm{H}} \\
		&=\mathbf{A}_{\mathrm{R}} \mathbf{\Lambda}\left(\mathbf{A}_{\mathrm{T}} \operatorname{diag}\left\{\left[ m_1, m_2, \cdots, m_L\right]\right\}\right)^{\mathrm{H}} \\
		&=\sum_{i=1}^{L} \tilde{\alpha}_{i} \mathbf{H}_{i},
	\end{aligned}
\end{equation}
where $\mathbf{H}_i=\boldsymbol{a}_{\mathrm{R},i}\boldsymbol{a}_{\mathrm{T},i}^{\mathrm{H}}$ denotes the $i$-th subchannel matrix, 
and $\tilde{\alpha}_i=\alpha_i m_i$ denotes the redistributed channel gain of the $i$-th scattering path which reveals the effect of the pattern. $(23)$ reveals that the effect of the pattern design is equivalent to the redistribution of {\color{black}gains} to all scattering paths. 

From the perspective of the singular value optimization, (10) can be transformed into {\color{black} \cite{34}}
\begin{equation}
	\begin{aligned}
		& -\log _{2} \operatorname{det}\left(\mathbf{I}_{N_{r}}+\frac{\rho}{N_{r}}\widetilde{\mathbf{H}}\widetilde{\mathbf{H}}^{\mathrm{H}}\right) \\
		=& -\log _{2} \prod_{i=1}^{r}\left(1+\frac{\rho}{N_{r}}\left|\sigma_i\left(\widetilde{\mathbf{H}}\right)\right|^2\right),
	\end{aligned}
\end{equation}
where $r=\operatorname{rank}\left(\widetilde{\mathbf{H}}\right)$ and $\sigma_i\left(\widetilde{\mathbf{H}}\right)$ denote the rank of $\widetilde{\mathbf{H}}$ and the $i$-th singular value of $\widetilde{\mathbf{H}}$, respectively. Based on (23) and (24), the pattern design problem in the single-pattern case can be generalized into:
\begin{equation}
	\begin{aligned}
		\mathcal{P}_{5}: \min _{\left\{x_i\right\}} \ &-\prod_{i=1}^{r}\left(1+{\color{black}\zeta} x_i\right) \\
		\text { s.t. } & \sum_{i=1}^{r}x_i-N_{r} N_{t} \leq 0, \\
		& {\color{black}\zeta}=\frac{\rho}{N_r},
	\end{aligned}
\end{equation}
where
\begin{equation}
	x_i=\left|\sigma_i\left(\widetilde{\mathbf{H}}\right)\right|^2=\left|\sigma_i\left(\sum_{l=1}^{L}\tilde{\alpha}_{l} \mathbf{H}_{l}\right)\right|^2.
\end{equation}

Without considering (26), the optimal solution $\left\{x_i^{\star}\right\}$ of $\mathcal{P}_5$ can be obtained easily through the Karush-Kuhn-Tucker (KKT) conditions. However, the non-convexity introduced by (26) generally excludes $\left\{x_i^{\star}\right\}$ from the feasible region.
\setcounter{equation}{\value{TempEqCnt}} 
\setcounter{TempEqCnt}{\value{equation}} 
\setcounter{equation}{32}                           
\begin{figure*}[hb]
	\hrulefill
	\begin{equation}
		\begin{aligned}
			\widehat{\mathbf{G}}_{i, j} &=\operatorname{Tr}\left(\widehat{\mathbf{H}}_{i}^{\mathrm{H}} \widehat{\mathbf{H}}_{j}\right) \\
			&=\operatorname{Tr}\left(\left(\boldsymbol{a}_{\mathrm{T}, i} \odot \widehat{\boldsymbol{m}}_{i}\right) \boldsymbol{a}_{\mathrm{R}, i}^{\mathrm{H}} \boldsymbol{a}_{\mathrm{R}, j}\left(\boldsymbol{a}_{\mathrm{T}, j} \odot \widehat{\boldsymbol{m}}_{j}\right)^{\mathrm{H}}\right) \\
			&=\frac{1}{N_r N_t}\sum_{n=1}^{N_{r}} e^{j 2 \pi \frac{d_{\mathrm{R}}}{\lambda}(n-1)\left(\sin \theta_{j}-\sin \theta_{i}\right)}  \sum_{k=1}^{N_{t}} \widehat{\boldsymbol{m}}_{i}(k) \widehat{\boldsymbol{m}}_{j}(k) e^{j 2 \pi \frac{d_{\mathrm{T}}}{\lambda}(k-1)\left(\sin \varphi_{i}-\sin \varphi_{j}\right)}.
		\end{aligned}
	\end{equation}
\end{figure*}
\setcounter{equation}{\value{TempEqCnt}} 

$(25)$ and $(26)$ exhibit a clear physical interpretation that the {\color{black}achievable rate} of the channel is equivalent to the sum {\color{black}achievable rate} of the independent eigen-subchannels. Each eigenvalue is the quantitative description of how much communication resource can be distributed to the corresponding eigen-subchannel. 
Similar to the principle of water-filling, a "good" channel means that the singular values of the channel matrix are distributed as even as possible, and vice verse. Based on that, we propose an eigenvalue optimization based {\color{black}gain} allocation scheme that aims to balance the singular values of $\widetilde{\mathbf{H}}$, i.e., minimizing the maximum singular value of $\widetilde{\mathbf{H}}$, given by
\setcounter{equation}{26}
\begin{equation}
	\begin{aligned}
		\mathcal{P}_6: \min_{\boldsymbol{p}} \ &   \sigma_{\mathrm{max}}\left(\widetilde{\mathbf{H}}\right) =\sigma_{\mathrm{max}}\left(\sum_{l=1}^{L}p_{l}\mathbf{H}_l\right) \\
		\text{ s.t. }& \sum_{l=1}^{L}p_{l}=1, \ p_{l}\geq0,
	\end{aligned}
\end{equation}
where $\boldsymbol{p}=\left[p_{1}, p_{2}, \ldots, p_{L}\right]^{\mathrm{T}}$ is the {\color{black} effective path gain vector. \footnote{\color{black}It is noted that the {\color{black}gain} allocation scheme aimed at maximizing the minimum singular value of $\widetilde{\mathbf{H}}$ has the same performance gain as minimizing the maximum singular value.} Without loss of generality, we consider the `min-max’ scheme in this paper.} $\mathcal{P}_6$ can be transformed into 
\begin{equation}
	\begin{aligned}
		\mathcal{P}_7: \min_{\boldsymbol{p}}\ & \lambda_{\mathrm{max}}\left(\widetilde{\mathbf{W}}\right) =\lambda_{\mathrm{max}}\left(\sum_{l=1}^{L}p_{l}\mathbf{W}_l\right) \\
		\text{ s.t. } & \sum_{l=1}^{L}p_{l}=1, \ p_{l}\geq0,
	\end{aligned}
\end{equation}
where $\widetilde{\mathbf{W}}=\left[\mathbf{0},\widetilde{\mathbf{H}};\widetilde{\mathbf{H}}^\mathrm{H},\mathbf{0}\right]$ and $\mathbf{W}_l=\left[\mathbf{0},\mathbf{H}_l;\mathbf{H}_l^\mathrm{H},\mathbf{0}\right]$ so that $\lambda_{\mathrm{max}}\left(\widetilde{\mathbf{W}}\right)=\sigma_{\mathrm{max}}\left(\widetilde{\mathbf{H}}\right)$ and $\lambda_{\mathrm{max}}\left(\mathbf{W}_l\right)=\sigma_{\mathrm{max}}\left(\mathbf{H}_l\right)$. $\lambda_{\mathrm{max}}\left(\widetilde{\mathbf{W}}\right)$ is the maximum eigenvalue of matrix $\widetilde{\mathbf{W}}$ {\color{black} \cite{38}}. {\color{black}With the upper bound $t$ of the maximum eigenvalue of $\widetilde{\mathbf{W}}$}, $\mathcal{P}_7$ can be further simplified as a SDP problem {\color{black} \cite{37}}:
\begin{equation}
	\begin{aligned}
		\mathcal{P}_8: \min_{\boldsymbol{p},t} \ & t \\
		\text{ s.t. }& \sum_{l=1}^L p_l \mathbf{W}_l-t\mathbf{I}_{N_t+N_r} \preceq 0, \\
		& \sum_{l=1}^{L}p_l=1, \ p_l\geq0.
	\end{aligned}
\end{equation}


$\mathcal{P}_8$ can be solved via CVX directly by expanding the first constraint into its equivalent real and imaginary representations. 
We introduce the power scaling factor $\delta$, given by (30) to ensure the satisfaction of the power constraint to the pattern channel matrix.
\begin{equation}
	\begin{aligned}
		&\left\|\delta\sum_{l=1}^{L}p_l\mathbf{H}_l\right\|_{\mathrm{F}}^2=N_tN_r \\
		\Leftrightarrow&{\delta}^2\mathrm{Tr}\left(\left(\sum_{l=1}^{L}p_l\mathbf{H}_l\right)^{\mathrm{H}}\left(\sum_{l=1}^{L}p_l\mathbf{H}_l\right)\right)=N_tN_r ,\\
		\Leftrightarrow&\delta=\sqrt{\frac{N_tN_r}{\mathrm{Tr}\left(\left(\sum_{l=1}^{L}p_l\mathbf{H}_l\right)^{\mathrm{H}}\left(\sum_{l=1}^{L}p_l\mathbf{H}_l\right)\right)}}.
	\end{aligned}
\end{equation}

Let $\left|\tilde{\alpha}_l\right|=p_l\delta$ for all $l=1,2,\ldots,L$. Then the corresponding $m_l$ of each scattering path is given by
\begin{equation}
	m_l=\frac{\left|\tilde{\alpha}_l\right|}{\left|\alpha_l\right|}=\frac{p_l\delta}{\left|\alpha_l\right|}.
\end{equation}
\subsection{Multi-Pattern Case}

When each antenna element employs a different pattern, $(16)$ can be transformed into
\begin{equation}
	\begin{aligned}
		\widetilde{\mathbf{H}} &=\mathbf{A}_{\mathrm{R}} \mathbf{\Lambda}\left(\mathbf{A}_{\mathrm{T}} \odot \mathbf{M}\right)^{\mathrm{H}} \\
		&=\sum_{i=1}^{L} \alpha_{i} \boldsymbol{a}_{\mathrm{R}, i}\left(\boldsymbol{a}_{\mathrm{T}, i} \odot {\color{black}\boldsymbol{m}_i}\right)^{\mathrm{H}} \\
		&=\sum_{i=1}^{L} \alpha_{i}p_i \boldsymbol{a}_{\mathrm{R}, i}\left(\boldsymbol{a}_{\mathrm{T}, i} \odot \widehat{\boldsymbol{m}}_{i}\right)^{\mathrm{H}}=\sum_{i=1}^{L} \alpha_{i}p_i\widehat{\mathbf{H}}_{i},
	\end{aligned}
\end{equation}
where $\widehat{\mathbf{H}}_{i}=\boldsymbol{a}_{\mathrm{R}, i}\left(\boldsymbol{a}_{\mathrm{T}, i} \odot \widehat{\boldsymbol{m}}_{i}\right)^{\mathrm{H}}$ is the $i$-th normalized modified subchannel with $\left\|\widehat{\mathbf{H}}_{i}\right\|_{\mathrm{F}}^2=1$ and $p_i$ is the {\color{black}gain} we distribute to the corresponding subchannel.  $\widehat{\boldsymbol{m}}_{i}=\frac{1}{p_i}\mathbf{M}(:,i)$ with $\left\|\widehat{\boldsymbol{m}}_{i}\right\|_2^2=N_t$ denotes the correlation modification vector of the $i$-th scattering path. {\color{black}$\boldsymbol{m}_i$ here denotes the $i$-th column vector of the matrix $\mathbf{M}$.}

\textit{Proposition 1:} The constraint $\left\|\widehat{\boldsymbol{m}}_{i}\right\|_2^2=N_t$ is equivalent to $\left\|\widehat{\mathbf{H}}_{i}\right\|_{\mathrm{F}}^2=1$.

{\it Proof:} See Appendix.

From $(32)$, we can observe the superiority of the multi-pattern case on the wireless channel for {\color{black}PR-MIMO} systems over the single-pattern case in Section IV-A. Compared with $(23)$, the introduction of $\widehat{\boldsymbol{m}}_{i}$ offers the ability to modify the subchannel matrix, which will change the correlation among each subchannel. Based on that, the system can redistribute {\color{black}channel gains} with a better correlation structure. 

In order to quantify the correlation structure of the pattern channel, we first present the covariance matrix $\widehat{\mathbf{G}} \in \mathbb{C}^{L \times L}$ of subchannels as (33) at the bottom of this page.
The $(i,j)$-th element of $\widehat{\mathbf{G}}$ describes the correlation of $\widehat{\mathbf{H}}_{i}$ and $\widehat{\mathbf{H}}_{j}$, and it is obvious that $\widehat{\mathbf{G}}$ is a Hermitian matrix. Based on $\widehat{\mathbf{G}}$, we define a correlation level indication vector in the following form:
\setcounter{equation}{33}
\begin{equation}
	\hat{\boldsymbol{g}}=\left[\sum_{j=2}^{L} \left|\widehat{\mathbf{G}}_{1,j}\right|^2, \sum_{j=1, j \neq 2}^{L} \left|\widehat{\mathbf{G}}_{2, j}\right|^2, \ldots, \sum_{j=1}^{L-1} \left|\widehat{\mathbf{G}}_{L, j}\right|^2\right]^{\mathrm{T}},
\end{equation}
which quantifies the sum correlation effect of each subchannel with all the other subchannels.  

Based on the above analysis, the efficient pattern design for the multi-pattern case can be decomposed into two steps. Firstly, we design the correlation modification matrix $\widehat{\mathbf{M}}=\left[\widehat{\boldsymbol{m}}_1,\widehat{\boldsymbol{m}}_2, \ldots, \widehat{\boldsymbol{m}}_L\right]$ to optimize the correlation structure of the pattern channel. Subsequently, based on the optimized correlation structure, we adopt the {\color{black}gain} allocation scheme introduced in Section IV-A to further improve the channel quality. In the following, we study the correlation modification problem.

Recall that the correlation coefficient between the $i$-th subchannel $\widehat{\mathbf{H}}_{i}$ and the $j$-th subchannel $\widehat{\mathbf{H}}_{j}$ is given by (33) at the bottom of this page. We can observe that the design of a single $\widehat{\boldsymbol{m}}_{i}$ will have an impact on the related $(L-1)$ correlation coefficients, which brings difficulties to decouple the design of each $\widehat{\boldsymbol{m}}_i$.

To deal with this problem, we adopt a sequential optimization framework, which sequentially updates the $L$ correlation modification vectors in an adaptive manner. The physical principle of the sequential optimization framework is to optimize the selected correlation vector in each iteration, so that the modified subchannel can maintain low correlation with all of the previously modified subchannels.
For each correlation modification vector, we formulate the optimization problem and present the algorithm based on manifold optimization and eigenvalue decomposition, as described below.

\subsubsection{Subproblem in SOF}
To begin with, the correlation modification matrix $\widehat{\mathbf{M}}$ is initialized as $\mathbf{1}_{M \times N}$, which means that there is no correlation modification for all subchannels. In the $i$-th iteration, we calculate the covariance matrix of each subchannel to obtain the correlation indication vector $\hat{\boldsymbol{g}}$ in (34). After sorting the entries of $\hat{\boldsymbol{g}}$ in a descending order, we can obtain the subchannel with the largest correlation level that is to be updated within the current iteration, whose index is denoted by $n_i$. Without loss of generality, considering the correlation coefficient between the $n_k$-th and the $n_i$-th normalized channel with $k \leq i-1$, $(33)$ can be simplified as
\begin{equation}
	\widehat{\mathbf{G}}_{n_{i}, n_{k}}=\rho_{n_{i}, n_{k}}^{\mathrm{R}} \boldsymbol{b}_{n_{i},n_{k}}^{\mathrm{T}} \widehat{\boldsymbol{m}}_{n_{i}},
\end{equation}
where 
\begin{equation}
	\boldsymbol{b}_{n_{i}, n_{k}}(n)=\frac{1}{N_t}\widehat{\boldsymbol{m}}_{n_{k}}(n) e^{j 2 \pi \frac{d_{\mathrm{T}}}{\lambda}(n-1)\left(\sin \varphi_{n_{i}}-\sin \varphi_{n_{k}}\right)}
\end{equation}
and $\rho_{n_{i}, n_{k}}^{\mathrm{R}}=\frac{1}{N_r}\sum_{n=1}^{N_{r}} e^{j 2 \pi \frac{d_{\mathrm{R}}}{\lambda}(n-1)\left(\sin \theta_{n_{k}}-\sin \theta_{n_{i}}\right)}$ denotes the correlation coefficient between the receive array response vectors of the corresponding subchannels. To minimize the correlation coefficient between the $n_k$-th modified subchannel and the $n_i$-th updated subchannel, we consider the following optimization problem:
\begin{equation}
	\begin{aligned}
		\mathcal{P}_{9}: \min _{\widehat{\boldsymbol{m}}_{n_{i}}} \ & \left|\widehat{\mathbf{G}}_{n_{i}, n_{k}}\right|^{2} \\
		\text { s.t. }& \widehat{\boldsymbol{m}}_{n_{i}}^{\mathrm{T}} \widehat{\boldsymbol{m}}_{n_{i}}=N_{t}, \\
		&\widehat{\boldsymbol{m}}_{n_{i}} \geq 0,
	\end{aligned}
\end{equation}
where
\begin{equation}
	\begin{aligned}
		\left|\widehat{\mathbf{G}}_{n_{i}, n_{k}}\right|^{2}&=\left|\rho_{n_{i}, n_{k}}^{\mathrm{R}}\right|^{2}\left(\boldsymbol{b}_{n_{i}, n_{k}}^{\mathrm{T}} \widehat{\boldsymbol{m}}_{n_{i}}\right)^{\mathrm{H}} \boldsymbol{b}_{n_{i}, n_{k}}^{\mathrm{T}} \widehat{\boldsymbol{m}}_{n_{i}} \\
		&=\left|\rho_{n_{i}, n_{k}}^{\mathrm{R}}\right|^{2} \widehat{\boldsymbol{m}}_{n_{i}}^{\mathrm{T}}\left(\boldsymbol{b}_{n_{i}, n_{k}}^{*} \boldsymbol{b}_{n_{i}, n_{k}}^{\mathrm{T}}\right) \widehat{\boldsymbol{m}}_{n_{i}}\\
		&=\widehat{\boldsymbol{m}}_{n_{i}}^{\mathrm{T}}\operatorname{real}\left\{\left|\rho_{n_{i}, n_{k}}^{\mathrm{R}}\right|^{2}\boldsymbol{b}_{n_{i}, n_{k}}^{*} \boldsymbol{b}_{n_{i}, n_{k}}^{\mathrm{T}}\right\}\widehat{\boldsymbol{m}}_{n_{i}}\\
		&=\widehat{\boldsymbol{m}}_{n_{i}}^{\mathrm{T}}\mathbf{B}_{n_{k}}\widehat{\boldsymbol{m}}_{n_{i}}
	\end{aligned}
\end{equation}
and the first constraint is the {\color{black}gain} constraint that ensures $\left\|\widehat{\mathbf{H}}_{n_{i}}\right\|_{\mathrm{F}}^{2}=1$. Considering that the real quadratic form will not be influenced by the imaginary component of the conjugate symmetric coefficient matrix, the cost function of $\mathcal{P}_{9}$ can be simplified by defining $\mathbf{B}_{n_{k}}=\operatorname{real}\left\{\left|\rho_{n_{i}, n_{k}}^{\mathrm{R}}\right|^{2}\boldsymbol{b}_{n_{i}, n_{k}}^{*} \boldsymbol{b}_{n_{i}, n_{k}}^{\mathrm{T}}\right\}$.

 Considering all the $(i-1)$ previously updated subchannels, the optimization problem in the $i$-th iteration is constructed as follows:
\begin{equation}
	\begin{aligned}
		\mathcal{P}_{10}: \min _{\widehat{\boldsymbol{m}}_{n_{i}}} \ &  \widehat{\boldsymbol{m}}_{n_{i}}^{\mathrm{T}}\left(\sum_{k=1}^{i-1} \mathbf{B}_{n_{k}}\right) \widehat{\boldsymbol{m}}_{n_{i}} \\
		\text { s.t. }& \widehat{\boldsymbol{m}}_{n_{i}}^{\mathrm{T}} \widehat{\boldsymbol{m}}_{n_{i}}=N_{t}, \\
		&\widehat{\boldsymbol{m}}_{n_{i}} \geq 0.
	\end{aligned}
\end{equation}

\subsubsection{Manifold Optimization based Algorithm for Solving (39)}

$\mathcal{P}_{10}$ is non-convex because of the quadratic equality constraint. Nevertheless, $\mathcal{P}_{10}$ can readily be solved using the conjugate gradient algorithm based on manifold optimization \cite{32}. The basic idea of manifold optimization is to handle the nonconvex problem in a Riemannian manifold space, and some optimization algorithms such as the conjugate gradient descend method can be analogously extended to the manifold space.

Defining the real circle as $\mathcal{M}=\left\{\mathbf{x} \in \mathbb{R}^{m}: \mathbf{x}^{\mathrm{T}} \mathbf{x}=1\right\}$, the gradient vector of the cost function at a given point $\mathbf{x} \in \mathcal{M}$ can be projected onto the tangent space $T_{\mathbf{x}} \mathcal{M}$
\begin{equation}
	\begin{aligned}
		\operatorname{grad} f(\mathbf{x}) &=\operatorname{Proj}_{\mathbf{x}} \nabla f(\mathbf{x}) \\
		&=\nabla f(\mathbf{x})-\nabla f(\mathbf{x}) \odot \mathbf{x}^2,
	\end{aligned}
\end{equation}
where $\nabla f(\mathbf{x})$ is the Euclidean gradient of the cost function. It can be verified that the Eculidean and Riemannian gradients of the objective function in (39) are
\begin{equation}
	\nabla f(\mathbf{x})=2\left(\sum_{k=1}^{i-1} \mathbf{B}_{n_{k}}\right)\mathbf{x}
\end{equation}
and 
\begin{equation}
	\operatorname{grad} f(\mathbf{x})=2\left(\sum_{k=1}^{i-1} \mathbf{B}_{n_{k}}\right)\mathbf{x} \odot \left(\mathbf{1}-\mathbf{x}^2\right).
\end{equation}

After determining the decent step using Armijo backtracking line search, we retract the descend vector to the manifold using the following operator
\begin{equation}
	\begin{aligned}
		\operatorname{Retr}_{\mathbf{x}}: & T_{\mathbf{x}} \mathcal{M} \rightarrow \mathcal{M}: \\
		& \alpha \mathbf{d} \mapsto \operatorname{Retr}_{\mathbf{x}}(\alpha \mathbf{d})=\operatorname{vec}\left[\frac{(\mathbf{x}+\alpha \mathbf{d})_{i}}{\left|(\mathbf{x}+\alpha \mathbf{d})_{i}\right|}\right],
	\end{aligned}
\end{equation}
where $\alpha$ denotes the step size and $\mathbf{d}$ denotes the descent direction in the tangent space. The overall algorithm is summarized in Algorithm 1, and the convergence of the algorithm can be guaranteed according to Theorem 4.3.1 in \cite{32,35}.
\begin{algorithm}[htb]
	\caption{Conjugate Gradient Algorithm Based on Manifold Optimization for Solving $\mathcal{P}_{10}$}
	\label{alg1}
	\begin{algorithmic}[1]
		\REQUIRE $\mathbf{B}=\sum_{k=1}^{i-1} \mathbf{B}_{n_{k}},\epsilon,T_{max}$
		\ENSURE $\widehat{\boldsymbol{m}}_{n_{i}}$
		\STATE Initialize $\boldsymbol{m}^{(0)}$ as $\boldsymbol{1}_{N_t \times 1}$ and set $k=1$;
		\STATE Calculate $\operatorname{grad} f \left(\boldsymbol{m}^{\left(0\right)}\right)$ via $(42)$ and obtain $\mathbf{d}^{(0)}=-\operatorname{grad} f \left(\boldsymbol{m}^{\left(0\right)}\right)$;
		\FOR{$\left|f\left(\boldsymbol{m}^{(k)}\right)-f\left(\boldsymbol{m}^{(k-1)}\right)\right| \geq \epsilon$ and $k \leq T_{max}$}
		\STATE Choose Armijo backtracking line search step size $\alpha^{(k)}$;
		\STATE Find the updated $\boldsymbol{m}^{\left(k+1\right)}$ via $(43)$: $\boldsymbol{m}^{\left(k+1\right)}=\operatorname{Retr}_{\boldsymbol{m}^{\left(k\right)}}\left(\alpha^{(k)}\mathbf{d}^{(k)}\right)$;
		\STATE Calculate gradient vector: $\mathbf{g}^{(k+1)}=\operatorname{grad} f\left(\boldsymbol{m}^{\left(k+1\right)}\right)$ via $(40)$-$(42)$;
		\STATE Choose Polak-Ribiere parameter $\beta^{(k+1)}=\frac{\operatorname{grad} f^{\mathrm{T}}\left(\boldsymbol{m}^{\left(k+1\right)}\right)\left(\operatorname{grad} f\left(\boldsymbol{m}^{\left(k+1\right)}\right)-\operatorname{grad} f\left(\boldsymbol{m}^{\left(k\right)}\right)\right)}{\left\|\operatorname{grad} f\left(\boldsymbol{m}^{\left(k\right)}\right)\right\|_2^2}$;
		
		\STATE Calculate conjugate direction $\mathbf{d}^{(k+1)}=-\mathbf{g}^{(k+1)}+\beta^{(k+1)}\mathbf{d}^{(k)}$;
		\STATE $k=k+1$;
		\ENDFOR
		\STATE Calculate the power scaling factor $\tau=\sqrt{\frac{N_t}{\left\|\max \left\{\boldsymbol{m}^{(k)},\boldsymbol{0}\right\}\right\|_2^2}}$ and obtain $\widehat{\boldsymbol{m}}_{n_i}^{\star}=\tau \max \left\{\boldsymbol{m}^{(k)},\boldsymbol{0}\right\}$;
	\end{algorithmic}
\end{algorithm}

\subsubsection{Eigenvalue Decomposition based Algorithm for Solving (39)}

Considering that the linear search process in the above Algorithm 1 might be computationally expensive, we propose a low-complexity alternative approach. To be more specific, based on the eigenvalue decomposition of $\mathbf{B}=\sum_{k=1}^{i-1} \mathbf{B}_{n_{k}}=\mathbf{U}\mathbf{\Sigma} \mathbf{U}^{\mathrm{T}}$, $\mathcal{P}_{10}$ can be simplified as
\begin{equation}
	\begin{aligned}
		\mathcal{P}_{11}: \min _{\boldsymbol{m}} \ & \left(\mathbf{U}^{\mathrm{T}}\boldsymbol{m}\right)^{\mathrm{T}} \mathbf{\Sigma} \left(\mathbf{U}^{\mathrm{T}}\boldsymbol{m}\right) \\
		\text { s.t. } & \boldsymbol{m}^{\mathrm{T}} \boldsymbol{m}=N_{t}, \\
		&\boldsymbol{m} \geq 0,
	\end{aligned}
\end{equation}
where $\mathbf{U}$ is an $N_t \times N_t$ real orthogonal matrix and $\mathbf{\Sigma}$ is the diagonal eigenvalue matrix. With $\boldsymbol{w}=\mathbf{U}^{\mathrm{T}}\boldsymbol{m}$, $\mathcal{P}_{11}$ can be transformed into $\mathcal{P}_{12}$:
\begin{equation}
	\begin{aligned}
		\mathcal{P}_{12}: \min _{\boldsymbol{w}} \ & \boldsymbol{w}^{\mathrm{T}} \mathbf{\Sigma} \boldsymbol{w} \\
		\text { s.t. } & \boldsymbol{w}^{\mathrm{T}} \boldsymbol{w}=N_{t}, \\
		&\mathbf{U}\boldsymbol{w} \geq 0.
	\end{aligned}
\end{equation}

Let $\boldsymbol{w}^{\star}=\sqrt{N_t} \mathbf{e}_{\mathrm{min}}$ where $\mathbf{e}_{\mathrm{min}}$ denotes a unit vector whose non-zero entry corresponds to the location of the smallest eigenvalue of $\mathbf{B}$ in $\operatorname{diag}\left\{\mathbf{\Lambda}\right\}$ \cite{34}. Therefore, a feasible solution to $\mathcal{P}_{11}$ can thus be obtained by
\begin{equation}
	\widehat{\boldsymbol{m}}_{n_{i}}^{\star}=\kappa\max \left\{\mathbf{U}\boldsymbol{w}^{\star}, \mathbf{0}\right\},
\end{equation}
where $\kappa=\sqrt{\frac{N_t}{\left\|\max \left\{\mathbf{U}\boldsymbol{w}^{\star}, \mathbf{0}\right\}\right\|_2^2}}$ is the power scaling factor to ensure the satisfaction of $\left\|\widehat{\boldsymbol{m}}_{n_{i}}^{\star}\right\|_2^2=N_t$.

Considering the non-convexity introduced by the quadratic equality constraint, it is difficult to obtain the optimal solution of $\mathcal{P}_{10}$. Fortunately, the mechanism of the SOF returns a point with a promising performance of the correlation modification process.
The overall algorithm for the sequential optimization framework is summarized in Algorithm 2.
\begin{algorithm}[htb]
	\caption{Sequential Optimization Framework (SOF)}
	\label{alg2}
	\begin{algorithmic}[1]
		\REQUIRE $\mathbf{A}_{\mathrm{R}}$, $\mathbf{A}_{\mathrm{T}}$, $\mathbf{\Lambda}$
		\ENSURE $\mathbf{M}$
		\STATE Initialize $\mathcal{I}=\emptyset$ and $\widehat{\mathbf{M}}=\mathbf{1}_{N_t \times L}$;
		\STATE Calculate $\mathbf{H}_i=\boldsymbol{a}_{\mathrm{R},i}\boldsymbol{a}_{\mathrm{T},i}^\mathrm{H}$; Obtain $\mathcal{S}=\left\{\mathbf{H}_{i} \mid i=1,2, \ldots, L\right\}$;
		\STATE Calculate $\widehat{\mathbf{G}}^{(1)}$ based on $(33)$; Calculate $\widehat{\boldsymbol{g}}^{(1)}$ based on $(34)$;
		\STATE Find $n_1$ such that $\widehat{\boldsymbol{g}}_{n_1}^{(1)}=\operatorname{max}\left\{\widehat{\boldsymbol{g}}^{(1)}\right\}$; Stack $\mathcal{I}=[\mathcal{I},n_1]$;
		\FOR{$i=2:L$}
		\STATE Remove $\widehat{\boldsymbol{g}}_j^{\left(i-1\right)}$ from $\widehat{\boldsymbol{g}}^{\left(i-1\right)}$ and obtain $\widehat{\boldsymbol{g}}'$, ${\forall}j \in \mathcal{I}$;
		\STATE Find $n_i$ such that $\widehat{\boldsymbol{g}}_{n_i}^{(i)}=\operatorname{max}\left\{\widehat{\boldsymbol{g}}'\right\}$; Stack $\mathcal{I}=\left[\mathcal{I},n_i\right]$;
		\STATE Initialize $\mathbf{B}_i=\boldsymbol{0}_{L \times L}$;
		\FOR{$k=1:(i-1)$}
		\STATE Obtain $n_k=\mathcal{I}(k)$;
		\STATE Calculate $\boldsymbol{b}_{n_i,n_k}$ based on $(36)$; 
		\STATE Obtain $\mathbf{B}_{n_k}=\boldsymbol{b}_{n_i,n_k}^{*}\boldsymbol{b}_{n_i,n_k}^{\mathrm{T}}$;
		\STATE Update $\mathbf{B}_i=\mathbf{B}_i+\mathbf{B}_{n_k}$;
		\ENDFOR
		\STATE Solve $\mathcal{P}_{10}$ and obtain $\widehat{\boldsymbol{m}}_{n_i}^{\star}$;
		\STATE Update $\widehat{\mathbf{M}}(:,n_i)=\widehat{\boldsymbol{m}}_{n_i}^{\star}$;
		\STATE Update $\widehat{\mathbf{H}}_{n_i}=\boldsymbol{a}_{\mathrm{R},n_i}\left(\boldsymbol{a}_{\mathrm{T},n_i} \odot \widehat{\mathbf{M}}(:,n_i)\right)^{\mathrm{H}}$;
		\STATE Calculate $\widehat{\mathbf{G}}^{(i)}$ based on $(33)$; Calculate $\widehat{\boldsymbol{g}}^{(i)}$ based on $(34)$;
		\ENDFOR
		\STATE Obtain $\widehat{\mathcal{S}}=\left\{\widehat{\mathbf{H}}_i \mid i=1,2, \ldots, L\right\}$;
		\STATE Calculate $\widehat{\mathbf{G}}$ based on $(33)$; Calculate $\widehat{\boldsymbol{g}}$ based on $(34)$;
		\STATE Solve $\mathcal{P}_8$ and obtain $\boldsymbol{p}$;
		\STATE Obtain $m_i$ based on $(31)$;
		\STATE Output $\mathbf{M}=\widehat{\mathbf{M}}\mathrm{diag}\left\{m_i\right\}$;
		
	\end{algorithmic}
\end{algorithm}

\section{Simulation Results}

Numerical results based on Monte Carlo simulations are presented in this section. We consider the multi-path channel model in $(2)$. Considering the scattering structure of the wireless environment, $(2)$ can be expanded into
\begin{equation}
	\mathbf{H}=\sum_{i=1}^{N_{\mathrm{cl}}}\sum_{j=1}^{N_{\mathrm{ray}}}\alpha_{i,j}\boldsymbol{a}_{\mathrm{R}}\left(\theta_{i,j}\right)\boldsymbol{a}_{\mathrm{T}}^{\mathrm{H}}\left(\varphi_{i,j}\right).
\end{equation}

The complex path gain $\alpha_{i,j}$ are $\text { i.i.d. } \mathcal{C} \mathcal{N}\left(0, \sigma_{i}^{2}\right)$, where $\sigma_{i}^{2}$ denotes the average power of the $i$-th cluster with $\sum_{i=1}^{N_{\mathrm{cl}}} \sigma_{i}^{2}=\gamma$, where $\gamma$ is the normalization parameter to ensure that $\mathbb{E}\left\{\left\|\mathbf{H}\right\|_{\mathrm{F}}^{2}\right\}=N_{r} N_{t}$. 
$\theta_{i,l}$ is uniformly distributed with mean $\theta_{i}$ and a standard deviation $\xi$, and $\varphi_{i,l}$ is uniformly distributed with mean $\varphi_{i}$ and the same standard deviation $\xi$. All the results are obtained by averaging over 1000 randomly generated channel realizations. Unless otherwise stated, $N_t=32$, $N_r=8$, $\xi=15^{\circ}$, and the half-wavelength antenna spacing is considered for both the transmitter and the receiver. Both $\theta_{i}$ and $\varphi_{i}$ are  uniformly distributed in the range of $\left[-\pi/2, \pi/2\right]$. {\color{black}In this section, the designed radiation patterns are presented to show the pattern reshaping process after reconfiguring operation. After that, the performance of the system {\color{black}achievable rate}} {\color{black}is shown to demonstrate the constructive effect of the designed pattern on the channel quality.}

\subsection{Radiation Pattern}
\begin{figure}[t]
	\centering
	
	\includegraphics[width=3.2in]{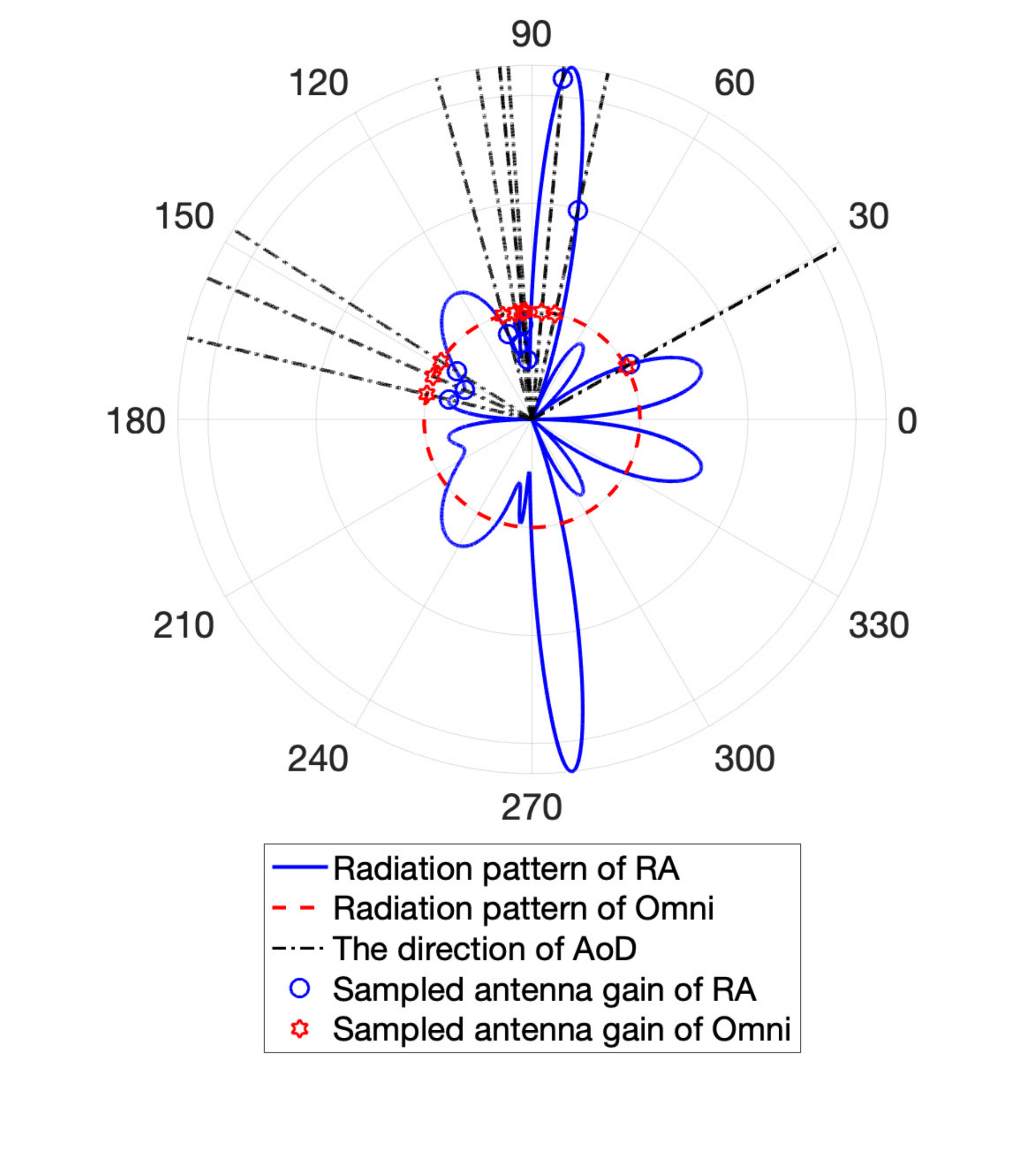}
	\caption{The radiation pattern for the single-pattern case, $N_t=32$,  $N_r=8$, $N_{\mathrm{cl}}=10$, $N_{\mathrm{ray}}=1$.}
	\label{structure}
\end{figure}
\begin{figure}[ht]
	\centering
	\subfloat[The radiation pattern of Transmit antenna 1]{\includegraphics[width=3.2in]{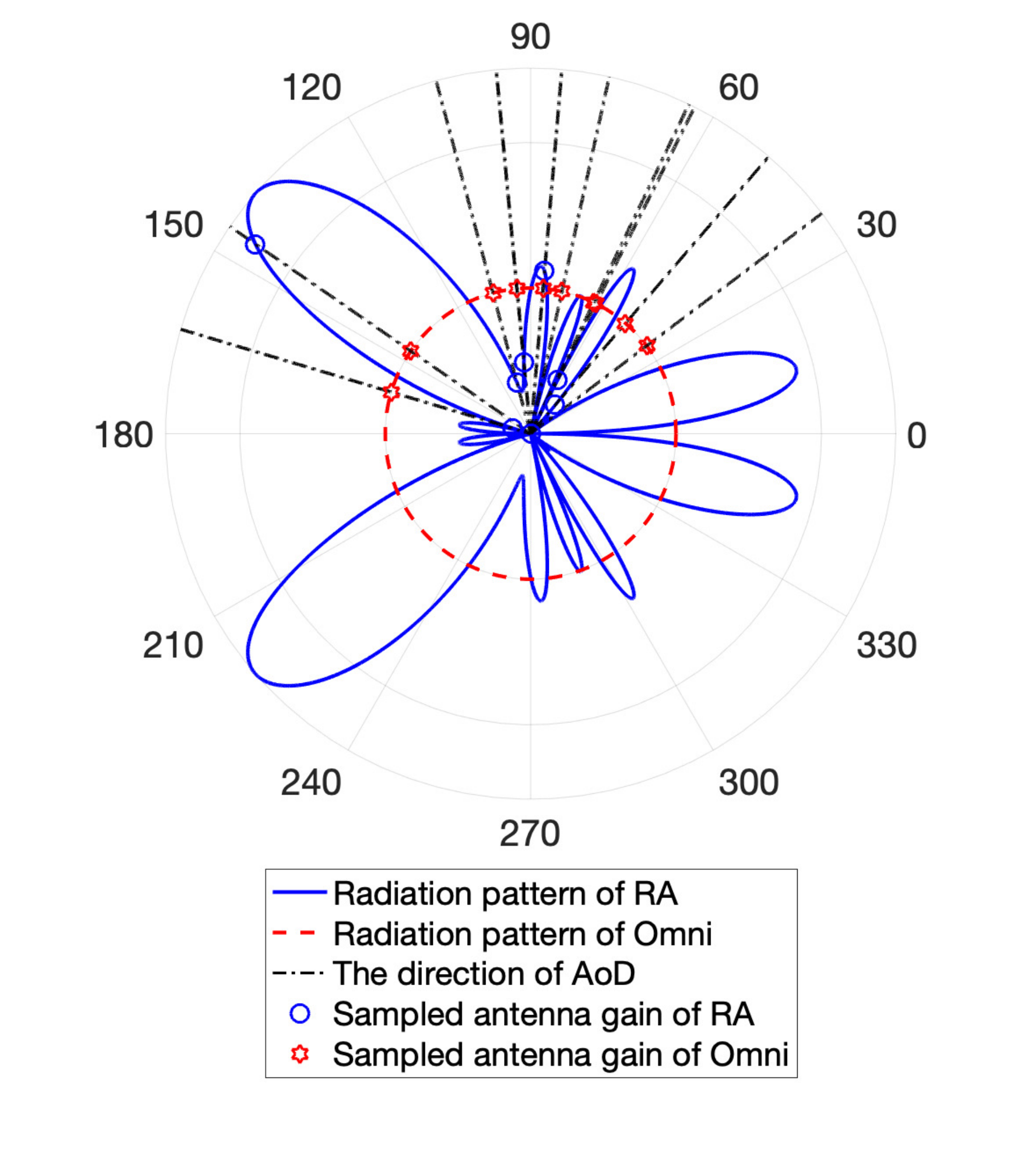}} 
	\hfill
	\subfloat[The radiation pattern of Transmit antenna 2]{\includegraphics[width=3.2in]{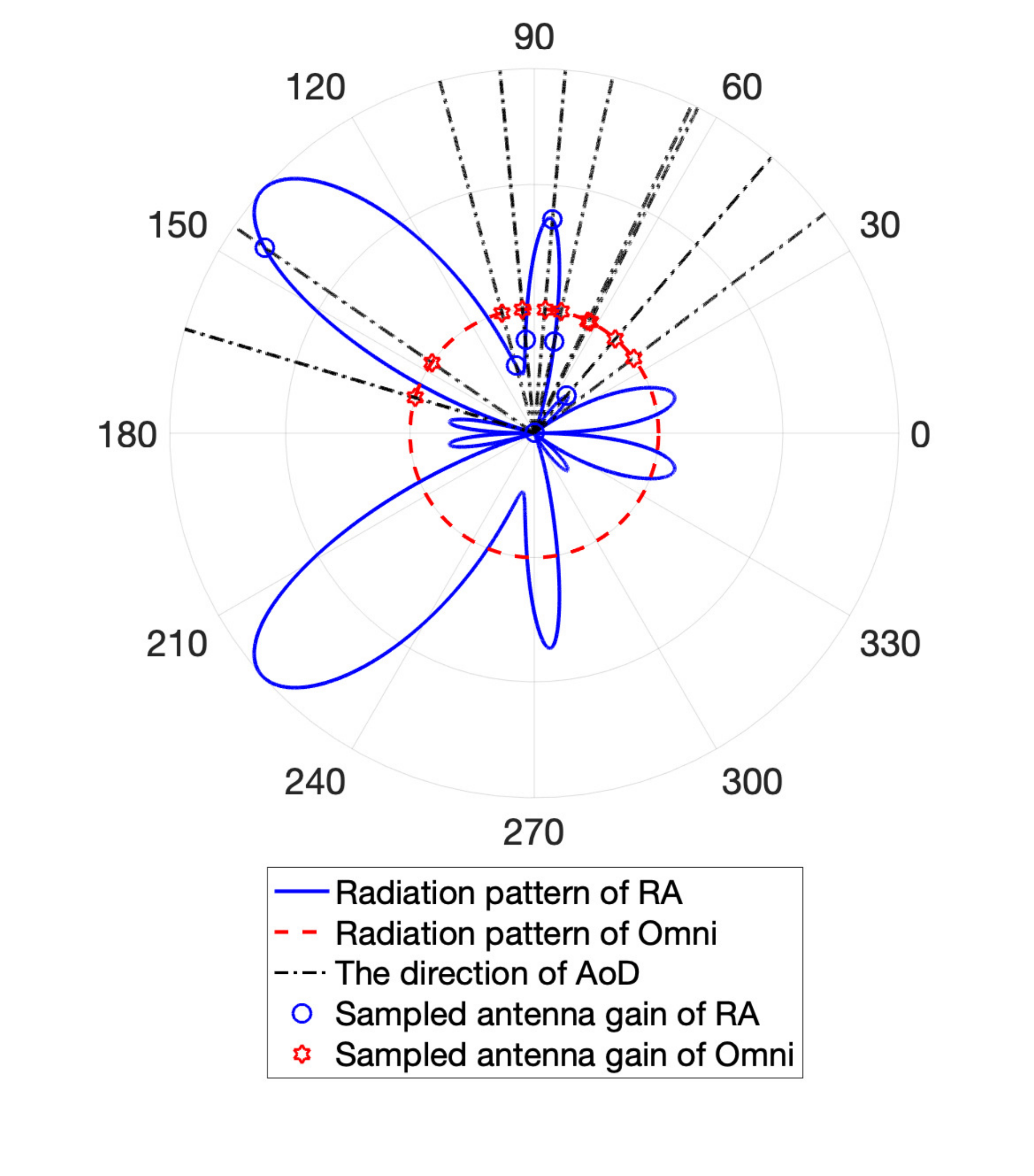}}
	\caption{The radiation pattern for the multi-pattern case, $N_t=32$,  $N_r=8$, $N_{\mathrm{cl}}=10$, $N_{\mathrm{ray}}=1$.}
\end{figure}

\begin{figure*}[htbp]
	\centering
	\includegraphics[width=7in]{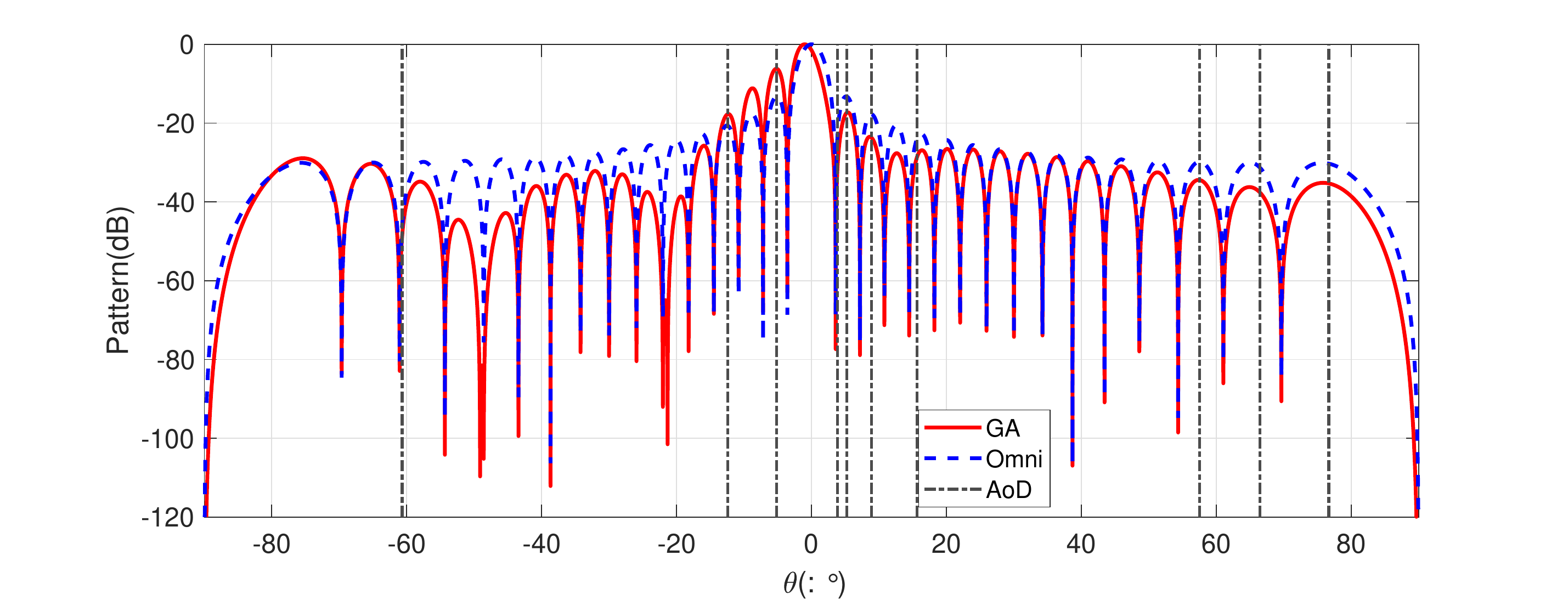}
	\caption{The array radiation pattern v.s. the scanning angle in the single-pattern case, $N_t=32$, $N_r=8$, $N_{\mathrm{cl}}=10$, $N_{\mathrm{ray}}=1$. }
	\label{structure}
\end{figure*}
\begin{figure*}[htbp]
	\centering
	\includegraphics[width=7in]{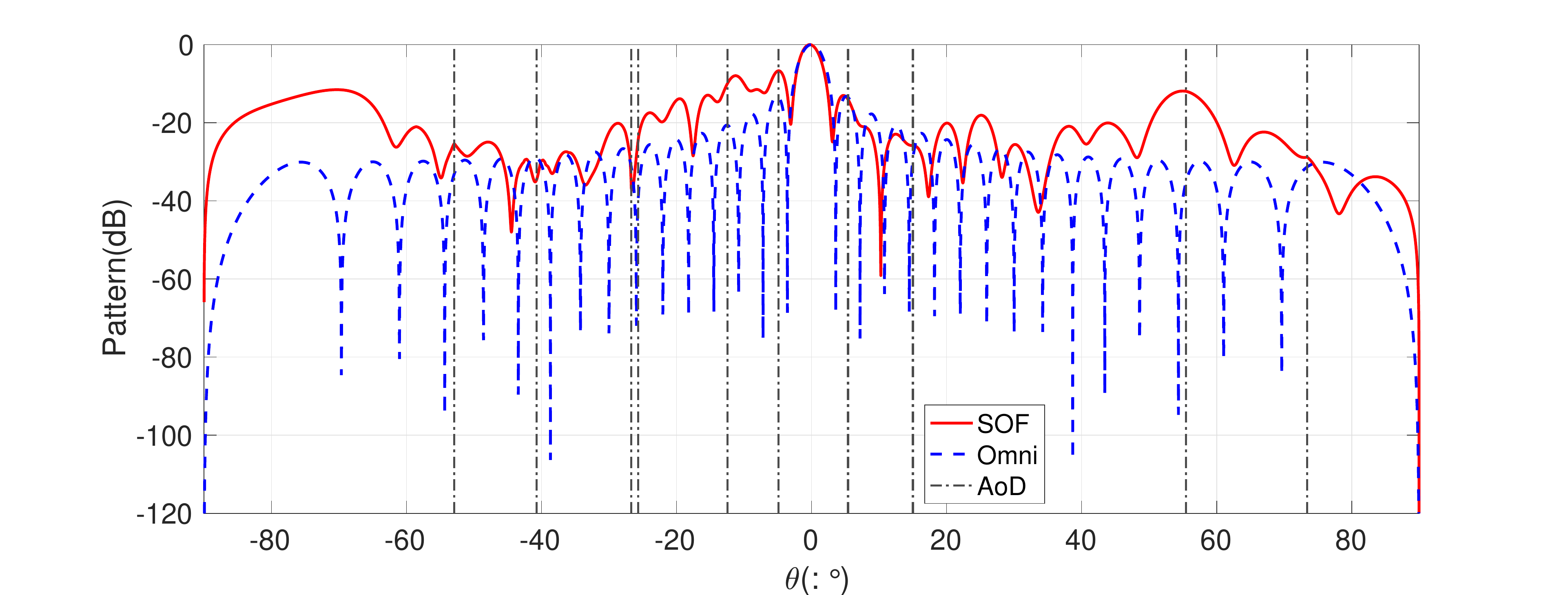}
	\caption{The array radiation pattern v.s. the scanning angle in the multi-pattern case, $N_t=32$, $N_r=8$, $N_{\mathrm{cl}}=10$, $N_{\mathrm{ray}}=1$. }
	\label{structure}
\end{figure*}

In this subsection, the radiation patterns for both of the single-pattern and the multi-pattern cases are presented via numerical interpolation based on the optimized pattern sampling matrices. {\color{black}We mainly discuss the radiation patterns of a representative antenna element to show the pattern reshaping process and the transmit antenna array to show the pattern effect on the array phase scanning.}

For convenience, the following abbreviations are used throughout this subsection:
\begin{enumerate}[]
	\item `RA': the designed reconfigurable antenna.
	\item `Omni': the omni antenna or the phased array equipped with omni antennas.
	\item `{\color{black}GA}': the array radiation pattern in the single-pattern system obtained by the {\color{black}gain} allocation ({\color{black}GA}) scheme.
	\item `SOF': the array radiation pattern in the multi-pattern system obtained by the sequential optimization framework (SOF) scheme.
\end{enumerate}

Through comparing the radiation patterns of reconfigurable and omni antennas in Fig. 2, we can realize the pattern reshaping after reconfigure operation. The sampling values of the pattern in the AoD directions are marked. The redistribution of the antenna gain is realized. We can observe that in a densely scattering region, the antenna gain will be limited for the high correlation, which is consistent with our design principle. In Fig. 3, we further present the radiation patterns of different RAs in the multi-pattern system. Comparing Fig. 3(a) and Fig. 3(b), we can observe that the correlation modification process results in different patterns on each antenna element. It is worth mentioning that the radiation pattern we designed only optimizes the antenna gain in the direction of AoD because the reconfigurable pattern affects the channel in the AoD directions when the pattern reconfigurability is only considered at the transmitter according to $(16)$. What's more, EoD and EoA default to $180^{\circ}$ to show the directionality of the resulting pattern.

Fig. 4 and Fig. 5 present the array radiation pattern of {\color{black}PR-MIMO} in the single-pattern and the multi-pattern cases, respectively. When each antenna element has the same pattern, the effect of the phase scanning still exists, compared with the traditional MIMO. {\color{black}The array radiation pattern of PR-MIMO reaches peaks and valleys synchronously compared with the traditional MIMO, revealing that} the influence of reconfigurable antennas focuses on the redistribution of the scanning power. From Fig. 5, we can observe that the phase scanning effect is destructed by the variety of patterns. What's more, the average power of scattering paths increases compared with the traditional MIMO for the ideal correlation structure obtained by SOF.

\subsection{{\color{black}Achievable Rate}}
The numerical results of the system {\color{black}achievable rate} are presented in this subsection. For the good-conditioned channel in the following part, $\left\{\sigma_{i}^{2}\right\}$ follows the normal distribution so that the channel condition number will be small. While for the ill-conditioned channel, we set $\sigma_{1}^{2}:\sigma_{2}^{2}:\sigma_{3}^{2}:\sigma_{4}^{2}: \ldots : \sigma_{N_{\mathrm{cl}}}^{2}=100:50:50:1: \ldots :1$ to obtain an extremely uneven distribution of $\left\{\sigma_{i}^{2}\right\}$ and the condition number of the ill-conditioned channel matrix is large.
The following abbreviations are used throughout this subsection:
\begin{enumerate}[]
	\item `Upper Bound': we adopt the ideal channel $\mathbf{H}_{\mathrm{opt}}=\sqrt{N_t}\mathbf{I}_{N_r \times N_t}$, considering that the eigenvectors of $\mathbf{H}_{\mathrm{opt}}$ have no influence on the {\color{black}achievable rate} calculation, according to $(10)$. Note that this upper bound is generally not achievable in practice. 
	\item `{\color{black}EOGA}': the proposed pattern design scheme in the single-pattern {\color{black}PR-MIMO} using the eigenvalue optimization based {\color{black}gain} allocation ({\color{black}EOGA}) scheme.
	\item `SDR': the suboptimal scheme in the single-pattern {\color{black}PR-MIMO} using the SDR algorithm through Gaussian approximation \cite{31}.
	\item `SOF-EVD-{\color{black}EOGA}': the proposed pattern design scheme in the multi-pattern {\color{black}PR-MIMO} using the eigenvalue decomposition-based (EVD-based) SOF with {\color{black}EOGA}.
	\item `SOF-MO-{\color{black}EOGA}': the proposed pattern design scheme in the multi-pattern {\color{black}PR-MIMO} using the manifold optimization-based (MO-based) SOF with {\color{black}EOGA}.
	\item `Good-Conditioned': the channel without the pattern reconfigurable antennas. The average power of each cluster has the same value.
	\item `Ill-Conditioned': the channel without the pattern reconfigurable antennas. The distribution of each cluster's average power is not even. 
\end{enumerate}

\begin{figure}[t]
	\centering
	
	\includegraphics[width=3.2in]{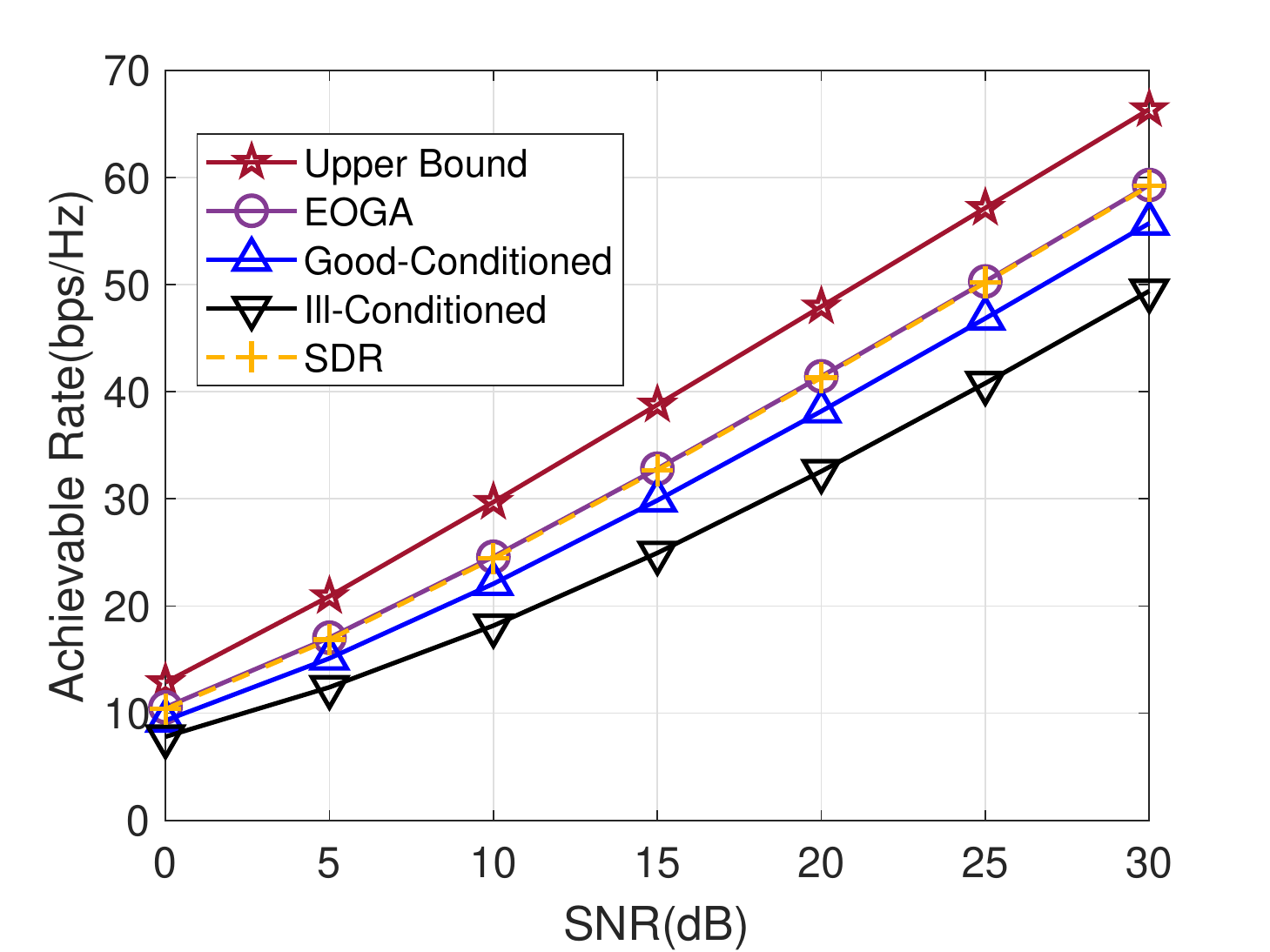}

	\caption{{\color{black}Achievable rate} v.s. transmit SNR, $N_t=32$, $N_r=8$, $N_{\mathrm{cl}}=10$, $N_{\mathrm{ray}}=8$, good-conditioned and ill-conditioned channels. }
	\label{structure}
\end{figure}
\begin{figure}[t]
	\centering
	\includegraphics[width=3.2in]{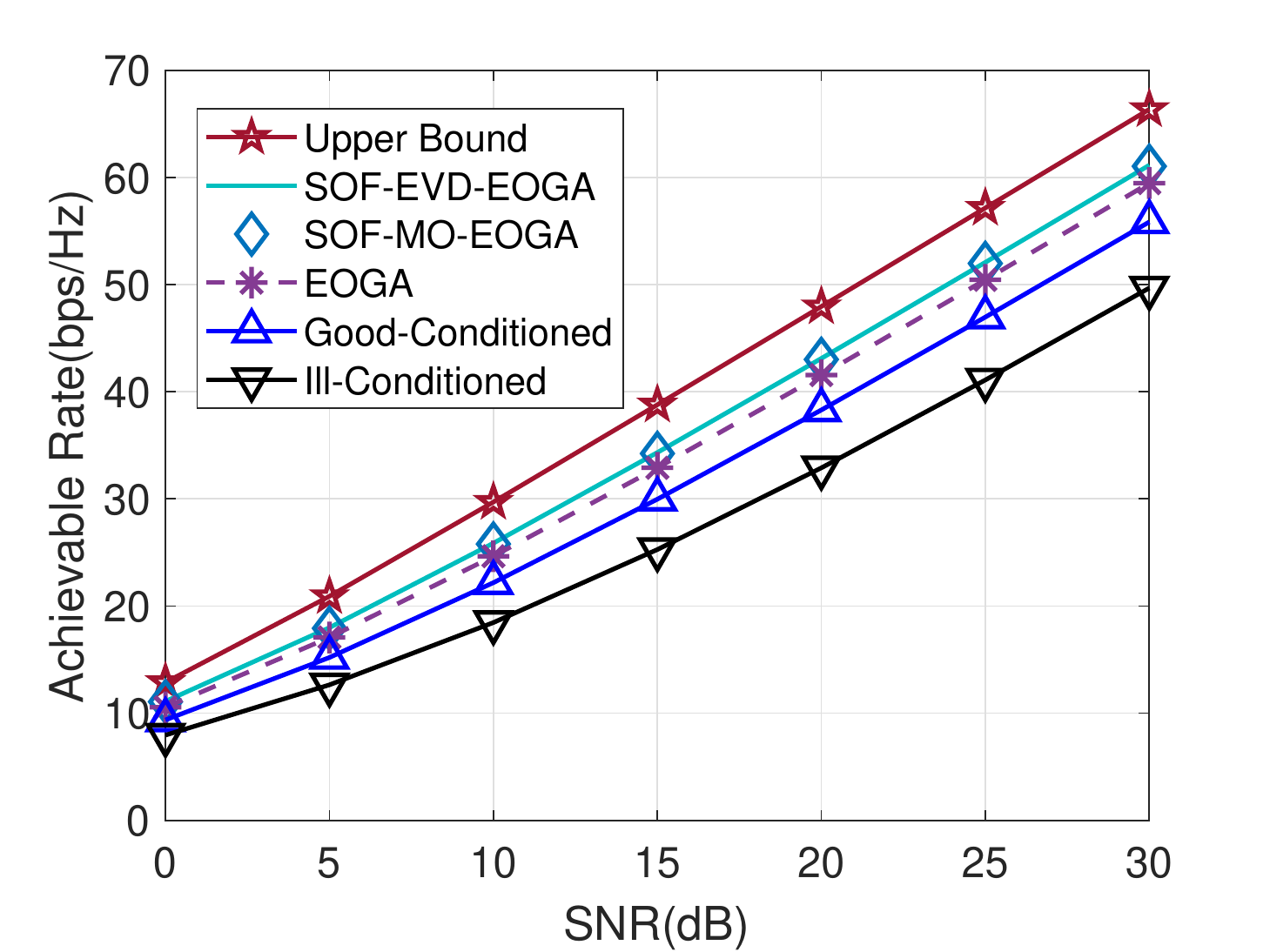}
	\caption{{\color{black}Achievable rate} v.s. transmit SNR, $N_t=32$, $N_r=8$, $N_{\mathrm{cl}}=10$, $N_{\mathrm{ray}}=8$, good-conditioned and ill-conditioned channels. }
	\label{structure}
\end{figure}
\begin{figure}[t]
	\centering
	\includegraphics[width=3.2in]{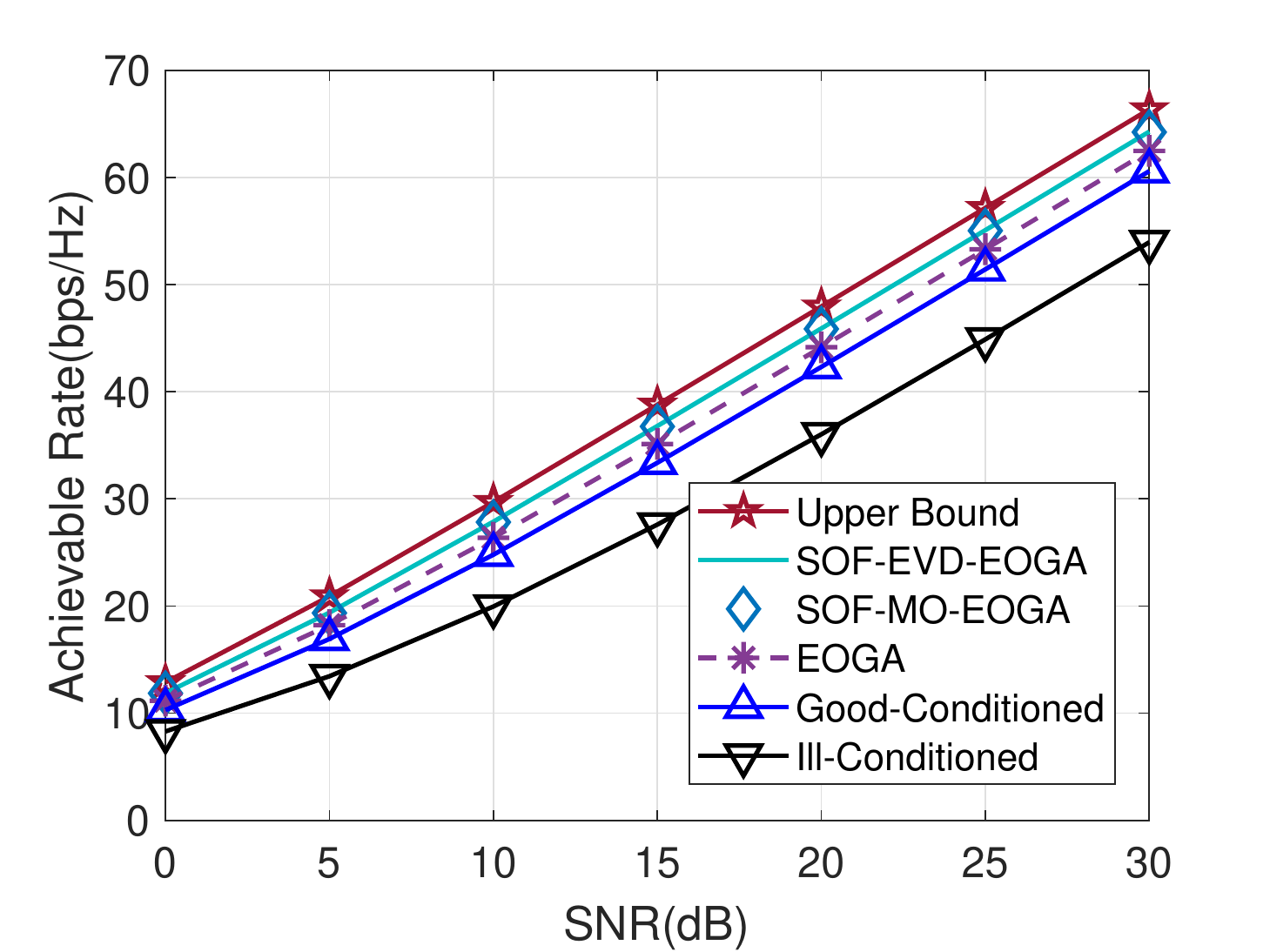}
	\caption{{\color{black}Achievable rate} v.s. transmit SNR, $N_t=32$, $N_r=8$, $N_{\mathrm{cl}}=20$, $N_{\mathrm{ray}}=8$, good-conditioned and ill-conditioned channels. }
	\label{structure}
\end{figure}
\begin{figure}[t]
	\centering
	\includegraphics[width=3.2in]{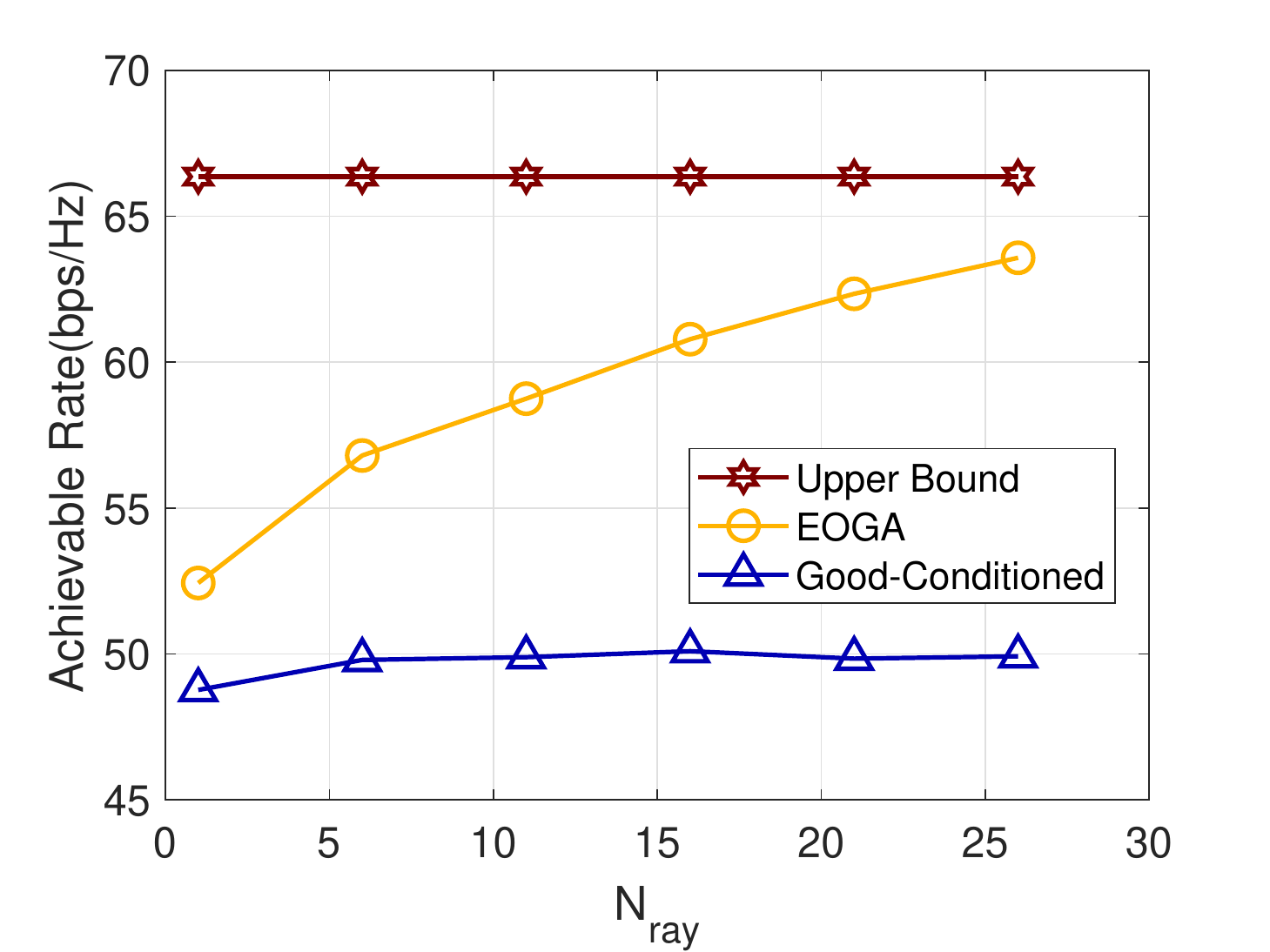}
	\caption{{\color{black}Achievable rate} v.s. number of rays in each cluster, $N_t=32$, $N_r=8$, $N_{\mathrm{cl}}=10$, $\xi=3^{\circ}$, SNR=30dB, good-conditioned channel. }
	\label{structure}
\end{figure}
In Fig. 6, the performance of {\color{black}EOGA} is shown compared with the upper bound with the ideal channel, the suboptimal solution obtained by the SDR with Gaussian approximation, and the physical channel equipped with omni-antennas both in good condition and in ill condition. We can see that the performance of {\color{black}EOGA} and SDR can provide a performance gain compared with the physical channel, which shows their superiority. Meanwhile, there is still a gap between the proposed pattern design schemes and the upper bound, which reveals that the lackness of phase modification ability limits the performance of the transmit reconfigurable pattern MIMO. 

In Fig. 7, the performance of SOF and {\color{black}EOGA} are presented compared with the upper bound, the physical channel in good condition and in ill condition when $N_{\mathrm{cl}}=10$. On one hand, in the single-pattern {\color{black}PR-MIMO}, {\color{black}EOGA} can provide a large performance gain especially when the physical channel is bad. However, {\color{black}EOGA} can not change the correlation structure of the channel, so the improvement compared with the good-conditioned channel is not apparent. On the other hand, in the multi-pattern {\color{black}PR-MIMO}, the quality of the channel can be significantly improved by the correlation modification so that SOF can further improve the performance of {\color{black}EOGA} and achieve a significant performance gain. The property is helpful in MIMO system design. With the pattern modification, the ill-conditioned channel will be transformed into good-conditioned.
 
Fig. 8 shows the analogous simulation results of Fig. 7 when $N_{\mathrm{cl}}=20$. Compared with Fig. 7, the {\color{black}achievable rate} of the physical channel in good condition is better for the additional subchannels. Moreover, the performance gap between the modified channel using SOF and the ideal channel becomes smaller, which shows that the modification ability is proportional to the number of clusters.  Meanwhile, more scattering paths provide more independent subpaths so that the gap between SOF and {\color{black}EOGA} becomes slighter. Finally, SOF-EVD-{\color{black}EOGA} and SOF-MO-{\color{black}EOGA} have similar performance.


Fig. 9 plots the performance of {\color{black}EOGA} in terms of the number of rays $N_{\mathrm{ray}}$ in each cluster
when $N_{\mathrm{cl}}=10$ and SNR=30dB. With the increasing of $N_{\mathrm{ray}}$, the performance of {\color{black}EOGA} is getting close to the upper bound, which reveals that the increased subchannels provide additional {\color{black}freedom for gain allocation} so that the {\color{black}gain} redistribution ability of {\color{black}EOGA} is enhanced. What's more, as the number of subchannels is large enough, the extra contribution due to {\color{black}EOGA} is getting marginal so that the slope is decreasing.


\section{Conclusions}

In this paper, we study the {\color{black}achievable rate} maximization pattern design for {\color{black}PR-MIMO} systems. We show that the effect of radiation reconfigurability can be regarded as an additional gain on the corresponding propagation directions. Based on that, the optimization problem of the optimal pattern design is formulated. We further discuss the single-pattern case where the optimized radiation pattern is the same for all the antenna elements, and the multi-pattern case where each antenna element can adopt different radiation patterns. More specifically, the pattern design in the single-pattern case is equivalent to a {\color{black}gain} redistribution among all scattering paths, and an eigenvalue optimization based {\color{black}gain} allocation algorithm is proposed. Compared with the single-pattern case, the multiple patterns in the multi-pattern case further offer the freedom of adjusting the correlation structure of the channel. A sequential optimization framework with manifold optimization and eigenvalue decomposition is proposed to obtain near-optimal solutions in the multi-pattern case. Numerical results validate the superiority of {\color{black}PR-MIMO} over traditional MIMO systems as well as the effectiveness of proposed algorithms. Our future work includes the application of symbol-level precoding in {\color{black}PR-MIMO} systems.




\appendix[Proof of Proposition 1]
Based on (32), the constraint $\left\|\widehat{\mathbf{H}}_i\right\|_{\mathrm{F}}^2$ can be transformed into
\begin{equation}
    \begin{aligned}
        \left\|\widehat{\mathbf{H}}_i\right\|_{\mathrm{F}}^2
        =&\operatorname{Tr}\left(\widehat{\mathbf{H}}_i^{\mathrm{H}}\widehat{\mathbf{H}}_i\right)\\
        =&\operatorname{Tr}\left(\left(\boldsymbol{a}_{\mathrm{T},i}\odot\widehat{\boldsymbol{m}}_i\right)\boldsymbol{a}_{\mathrm{R},i}^{\mathrm{H}}\boldsymbol{a}_{\mathrm{R},i}\left(\boldsymbol{a}_{\mathrm{T},i}\odot\widehat{\boldsymbol{m}}_i\right)^{\mathrm{H}}\right)\\
        =&\frac{1}{N_t}\sum_{j=1}^{N_t}\widehat{m}_{i,j}^2=\frac{1}{N_t}\left\|\widehat{\boldsymbol{m}}_i\right\|_2^2,
    \end{aligned}
\end{equation}
where $\boldsymbol{a}_{\mathrm{R},i}^{\mathrm{H}}\boldsymbol{a}_{\mathrm{R},i}=1$ and $\widehat{m}_{i,j}$ denotes the $j$-th element of $\widehat{\boldsymbol{m}}_i$. Based on (48), the equivalence between $\left\|\widehat{\mathbf{H}}_i\right\|_{\mathrm{F}}^2=1$ and $\left\|\widehat{\boldsymbol{m}}_i\right\|_2^2=N_t$ is proved.
\bibliographystyle{IEEEtran}
\bibliography{IEEEreference}

\begin{IEEEbiography}[{\includegraphics[width=1in,height=1.25in,clip,keepaspectratio]{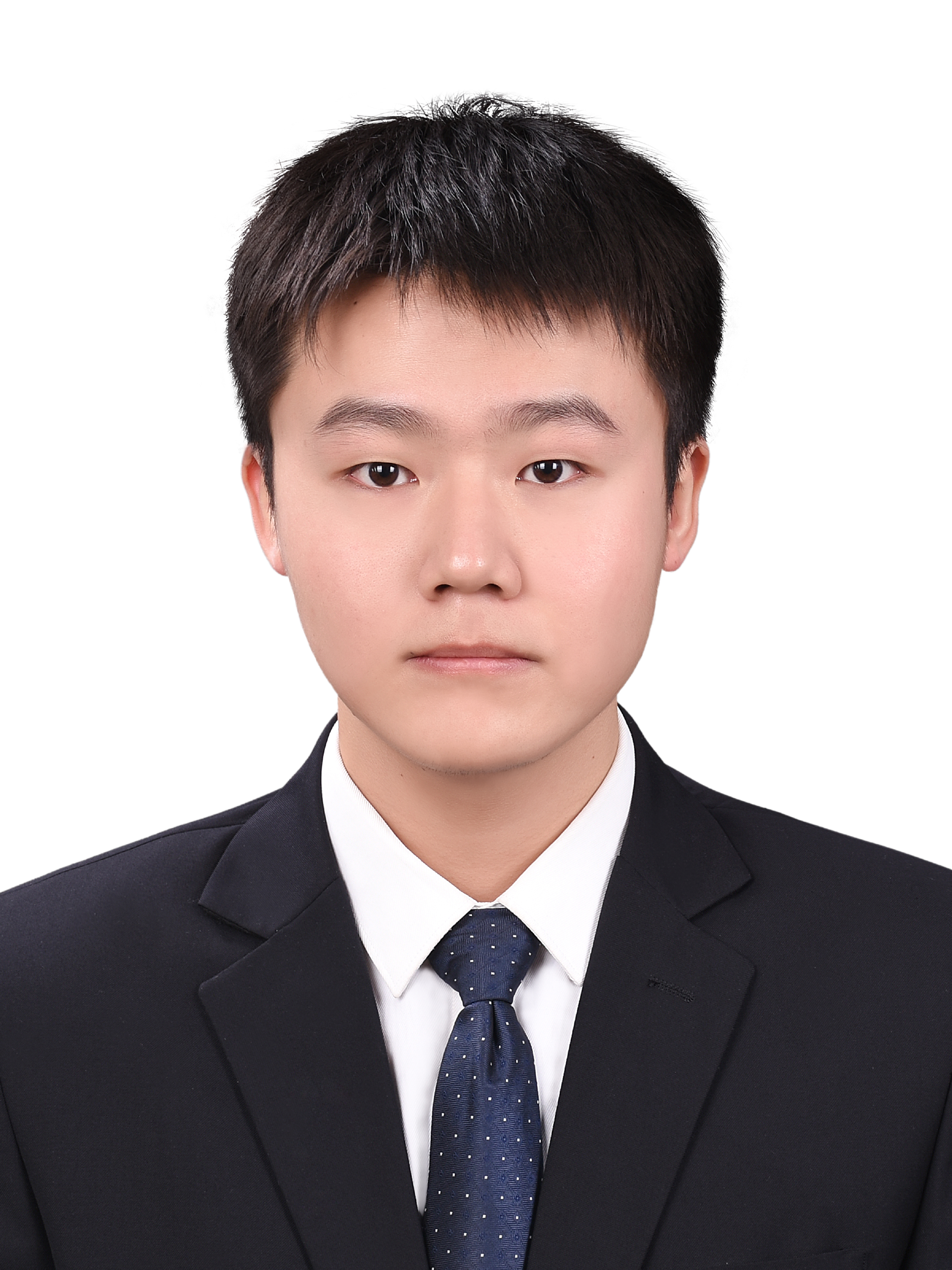}}]{Haonan Wang}
 (S’22) received the bachelor’s degree in information engineering from Xi’an Jiaotong University in 2020. He is currently pursuing the master’s degree with the School of Information and Communications Engineering, Faculty of Electronic and Information Engineering, Xi’an Jiaotong University. His research interests lie in the physical-layer techniques for wireless communications, including Reconfigurable MIMO, Symbol-Level Precoding, and Convex Optimization.
\end{IEEEbiography}
\begin{IEEEbiography}[{\includegraphics[width=1in,height=1.25in,clip,keepaspectratio]{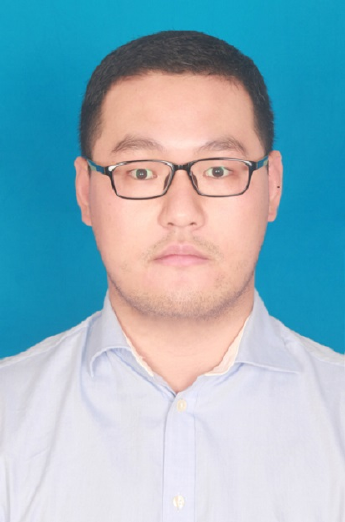}}]{Ang Li} (S'14-M'18-SM'21) received his Ph.D. degree in the Communications and Information Systems research group, Department of Electrical and Electronic Engineering, University College London in April 2018. He was a postdoctoral research associate in the School of Electrical and Information Engineering, The University of Sydney from May 2018 to February 2020. He joined Xi’an Jiaotong University in March 2020 and is now a Professor in the School of Information and Communications Engineering, Faculty of Electronic and Information Engineering, Xi'an Jiaotong University, Xi'an, China. His main research interests lie in the physical-layer techniques in wireless communications, including MIMO/massive MIMO, interference exploitation, symbol-level precoding, and reconfigurable MIMO, etc. He currently serves as the Associate Editor for IEEE Communications Letters, IEEE Open Journal of Signal Processing, and EURASIP Journal on Wireless Communications and Networking. He is the recipient of the 2021 IEEE Signal Processing Society Young Author Best Paper Award. He has been an Exemplary Reviewer for IEEE Communications Letters, IEEE Transactions on Communications, and IEEE Wireless Communications Letters. He has served as the Co-Chair of the IEEE ICASSP 2020 Special Session on 'Hardware-Efficient Large-Scale Antenna Arrays: The Stage for Symbol-Level Precoding', and has organized a Tutorial in IEEE ICC 2021 on 'Interference Exploitation through Symbol Level Precoding: Energy Efficient Transmission for 6G and Beyond'.
\end{IEEEbiography}
\begin{IEEEbiography}[{\includegraphics[width=1in,height=1.25in,clip,keepaspectratio]{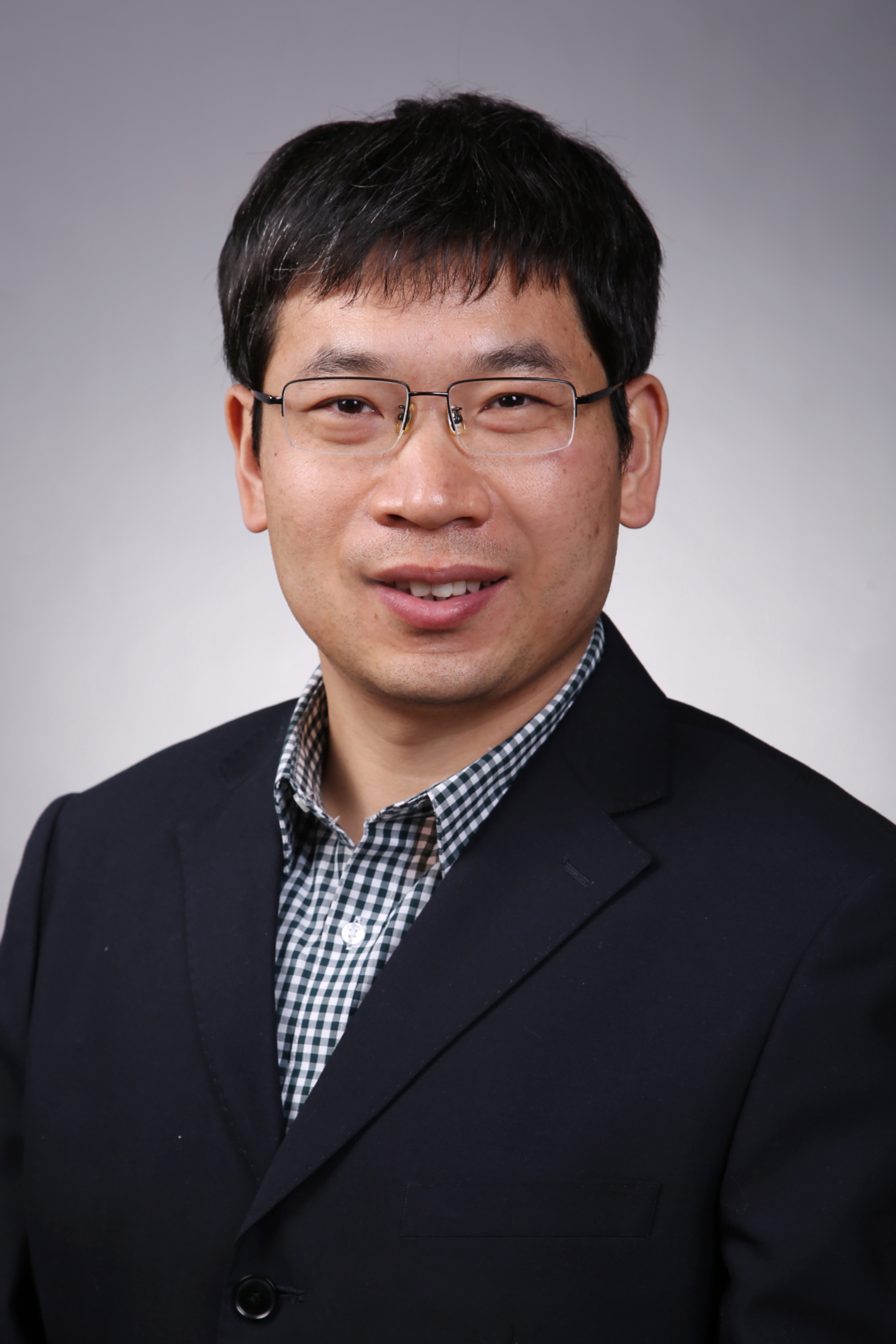}}]{Ya-Feng Liu} (M'12--SM'18) received the B.Sc. degree in applied mathematics from Xidian University, Xi'an, China, in 2007, and the Ph.D. degree in computational mathematics from the Chinese Academy of Sciences (CAS), Beijing, China, in 2012. During his Ph.D. study, he was supported by the Academy of Mathematics and Systems Science (AMSS), CAS, to visit Professor Zhi-Quan (Tom) Luo at the University of Minnesota (Twins Cities) from 2011 to 2012. After his graduation, he joined the Institute of Computational Mathematics and Scientific/Engineering Computing, AMSS, CAS, Beijing, China, in 2012, where he became an Associate Professor in 2018. His main research interests are nonlinear optimization and its applications to signal processing, wireless communications, and machine learning. 

Dr. Liu currently serves as an Associate Editor for the IEEE Transactions on Signal Processing, the IEEE Signal Processing Letters, and the Journal of Global Optimization. He served as an Editor for the IEEE Transactions on Wireless Communications (2019--2022). He is an elected member of the Signal Processing for Communications and Networking Technical Committee (SPCOM-TC) of the IEEE Signal Processing Society (2020--2022 and 2023--2025). He received the Best Paper Award from the IEEE International Conference on Communications (ICC) in 2011, the Chen Jingrun Star Award from the AMSS in 2018, the Science and Technology Award for Young Scholars from the Operations Research Society of China in 2018, the 15th IEEE ComSoc Asia-Pacific Outstanding Young Researcher Award in 2020, and the Science and Technology Award for Young Scholars from China Society for Industrial and Applied Mathematics in 2022. Students supervised and co-supervised by him won the Best Student Paper Award from the International Symposium on Modeling and Optimization in Mobile, Ad Hoc and Wireless Networks (WiOpt) in 2015 and the Best Student Paper Award of IEEE International Conference on Acoustics, Speech and Signal Processing (ICASSP) in 2022.
\end{IEEEbiography}


\begin{IEEEbiography}[{\includegraphics[width=1in,height=1.25in,clip,keepaspectratio]{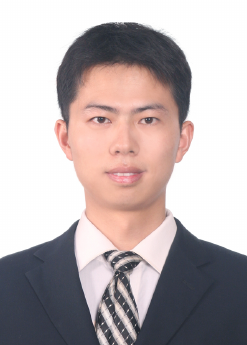}}]{Qibo Qin} received the B.S. degree in Electronic and Information Engineering from Xian Jiao Tong University, Xian, China, in 2014, and the Ph.D. degree from Shanghai Jiao Tong University, Shanghai, China, in 2019. He is currently an algorithm engineer in Huawei Technologies Ltd., Shanghai, China. His current research interests include wireless channel measurement and modeling, massive multiple-input multiple-output systems, digital twin and millimeter wave communications.
\end{IEEEbiography}
\begin{IEEEbiography}[{\includegraphics[width=1in,height=1.25in,clip,keepaspectratio]{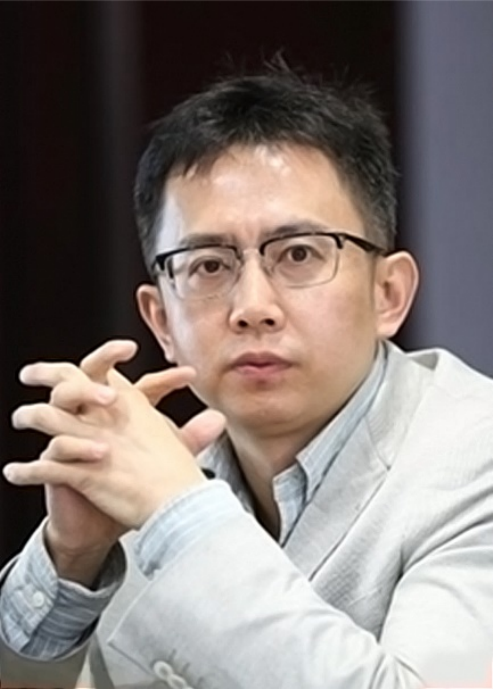}}]{Lingyang Song} (S’03-M’06-SM’12-F'19) received his PhD from the University of York, UK, in 2007, where he received the K. M. Stott Prize for excellent research. He worked as a research fellow at the University of Oslo, Norway until rejoining Philips Research UK in March 2008. In May 2009, he joined the School of Electronics Engineering and Computer Science, Peking University, and is now a Boya Distinguished Professor. His main research interests include wireless communications, mobile computing, and machine learning.

Dr. Song has co-authored of 2 text books, including “Wireless Device-to- Device Communications and Networks,” and “Full-Duplex Communications and Networks,” by Cambridge University Press, UK. He is the co-author of 17 best paper awards and 1 best demo award, including IEEE Leonard G. Abraham Prize in 2016, IEEE Communications Society Heinrich Hertz Award in 2021, and IEEE Communication Society Asia-Pacific Outstanding Paper Award in 2021, best paper awards from IEEE Communication Society Flagship Conference: IEEE ICC 2014, IEEE ICC 2015, IEEE Globecom 2014, and the best demo award in the ACM Mobihoc 2015. 
 
Dr. Song has been elected to serve the IEEE Vehicular Technology Society Board of Governors (2022-2024). He has served as a Distinguished Lecturer of IEEE Communications Society (2015-2018), an Area Editor of IEEE Transactions on Vehicular Technology (2019-), an Editor of IEEE Transactions on Communications (2019-), an Editor of China Communications (2015-), and an Editor of Transactions on Wireless Communications (2013-2018). He also serves as a Section Editor for Springer Handbook of Cognitive Radio (2016-). He served as the TPC co-chairs for ICUFN 2011/2012 and IEEE ICCC 2019. He served as symposium co-chairs for IEEE ICC 2014/2016, IEEE VTC 2016 spring, and IEEE Globecom 2016. He has served as Vice Chair (2016-2019) and Chair (2021-2022) of IEEE Communications Society Cognitive Network Technical Committee, and Vice Chair (2016-2019) and Chair (2020-2021) of IEEE Communications Society Asia Pacific Board Technical Affairs Committee.
\end{IEEEbiography}
\begin{IEEEbiography}[{\includegraphics[width=1in,height=1.25in,clip,keepaspectratio]{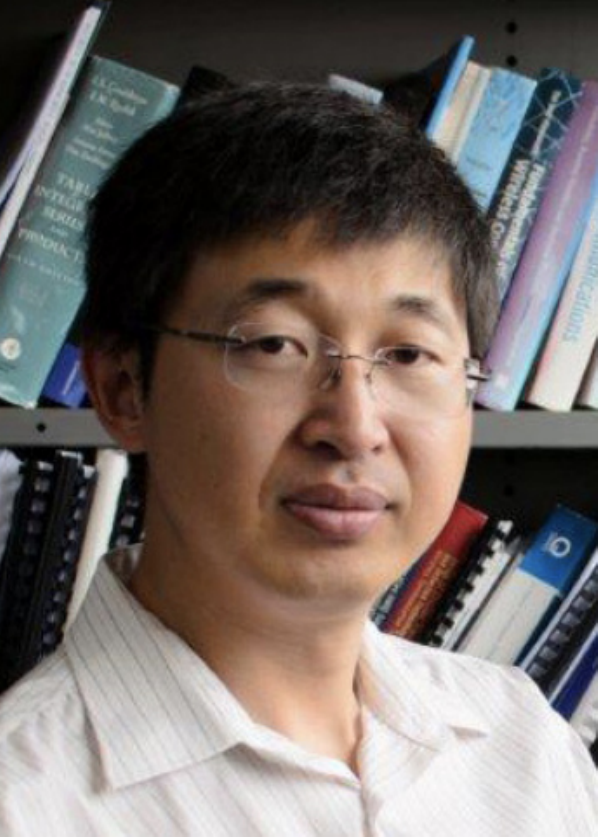}}]{Yonghui Li} (M’04-SM’09-F’19) received his PhD degree in November 2002 from Beijing University of Aeronautics and Astronautics. From 1999 – 2003, he was affiliated with Linkair Communication Inc, where he held a position of project manager with responsibility for the design of physical layer solutions for the LAS-CDMA system. Since 2003, he has been with the Centre of Excellence in Telecommunications, the University of Sydney, Australia. He is now a Professor in School of Electrical and Information Engineering, University of Sydney. He is the recipient of the Australian Queen Elizabeth II Fellowship in 2008 and the Australian Future Fellowship in 2012. His current research interests are in the area of wireless communications, with a particular focus on MIMO, millimeter wave communications, machine to machine communications, coding techniques and cooperative communications. He holds a number of patents granted and pending in these fields. He is now an editor for IEEE transactions on communications and IEEE transactions on vehicular technology. He also served as a guest editor for several special issues of IEEE journals, such as IEEE JSAC special issue on Millimeter Wave Communications. He received the best paper awards from IEEE International Conference on Communications (ICC) 2014, IEEE PIMRC 2017 and IEEE Wireless Days Conferences (WD) 2014. He is Fellow of IEEE.
\end{IEEEbiography}
\end{document}